\documentclass[a4paper]{scrartcl}

\usepackage[greek,american]{babel}
\usepackage[T1]{fontenc}
\usepackage[latin1]{inputenc}

\usepackage{lmodern}

\usepackage{amsmath}
\usepackage{amssymb,amsfonts,textcomp}
\usepackage{color}

\usepackage{hyperref}
\hypersetup{pdftex, colorlinks=true, linkcolor=blue, citecolor=blue, filecolor=blue, urlcolor=blue, pdftitle=Exoplanet Clouds Chapter v2.1.docx}
\usepackage{graphicx}

\newcommand\textstyleInternetlink[1]{\textcolor[rgb]{0.0,0.0,0.5019608}{#1}}
\title{Exoplanet Clouds Chapter v2.1.docx}

\begin{document}
\clearpage\setcounter{page}{1}

\begin{center}
\textbf{\huge{Clouds and Hazes in Exoplanet\\ Atmospheres}}
\end{center}

\bigskip
\bigskip

\begin{center}
\begin{Large}
Mark S. Marley\\
\textit{NASA Ames Research Center}

\bigskip

Andrew S. Ackerman
\par
\textit{NASA Goddard Institute for Space Studies}

\bigskip

Jeffrey N. Cuzzi
\par
\textit{NASA Ames Research Center}

\bigskip

Daniel Kitzmann
\par
\textit{Zentrum f\"ur Astronomie und Astrophysik,\\ Technische Universit\"at Berlin}
\end{Large}
\end{center}

\clearpage

\section*{Abstract}

Clouds and hazes are commonplace in the atmospheres of solar system planets and are likely ubiquitous in the atmospheres of extrasolar planets as well. Clouds affect every aspect of a planetary atmosphere, from the transport of radiation, to atmospheric chemistry, to dynamics and they influence - if not control - aspects such as surface temperature and habitability. In this review we aim to provide an introduction to the role and properties of clouds in exoplanetary atmospheres. We consider the role clouds play in influencing the spectra of planets as well as their habitability and detectability. We briefly summarize how clouds are treated in terrestrial climate models and consider the far simpler approaches that have been taken so far to model exoplanet clouds, the evidence for which we also review. Since clouds play a major role in the atmospheres of certain classes of brown dwarfs we briefly discuss brown dwarf cloud modeling as well. We also review how the scattering and extinction efficiencies of cloud particles may be approximated in certain limiting cases of small and large particles in order to facilitate physical understanding. Since clouds play such important roles in planetary atmospheres, cloud modeling may well prove to be the limiting factor in our ability to interpret future observations of extrasolar planets.

\section{Introduction}

Clouds and hazes are found in every substantial solar system atmosphere and are likely ubiquitous in extrasolar planetary atmospheres as well. They provide sinks for volatile compounds and influence both the deposition of incident flux and the propagation of emitted thermal radiation. Consequently they affect the atmospheric thermal profile, the global climate, the spectra of scattered and emitted radiation, and the detectability by direct imaging of a planet. As other chapters in this book attest, clouds and hazes are a complex and deep subject. In this chapter we will broadly discuss the roles of clouds and hazes as they relate to the study of exoplanet atmospheres. In particular we will focus on the challenge of exoplanet cloud modelling and discuss the impact of condensates on planetary climates.

The terms ``clouds'' and ``hazes'' are sometimes used interchangeably. Here we use the term ``cloud'' to refer to condensates that grow from an atmospheric constituent when the partial pressure of the vapor exceeds its saturation vapor pressure. Such supersaturation is typically produced by atmospheric cooling, and cloud particles will generally evaporate or sublimate in unsaturated conditions. A general framework for such clouds in planetary atmospheres is provided by \textit{S\'anchez-Lavega et al.} (2004). By ``haze'' we refer to condensates of vapor produced by photochemistry or other non-equilibrium chemical processes. This usage is quite different from that of the terrestrial water cloud microphysics literature where the distinction depends on water droplet size and atmospheric conditions.

Because exoplanetary atmospheres can plausibly span such a wide range of compositions as well as temperature and pressure conditions, a large number of species may form clouds. Depending on conditions, clouds in a solar composition atmosphere can include exotic refractory species such as Al\textsubscript{2}O\textsubscript{3}, CaTiO\textsubscript{3}, Mg\textsubscript{2}SiO\textsubscript{4} and Fe at high temperature and Na\textsubscript{2}S, MnS, and of course H\textsubscript{2}O at lower temperatures. Many other species condense as well including CO\textsubscript{2} in cold, Mars-like atmospheres and NH\textsubscript{3} in the atmospheres of cool giants, like Jupiter and Saturn. In Earth-like atmospheres water clouds are likely important although Venus-like conditions and sulfuric-acid or other clouds are possibilities as well. Depending on atmospheric temperature, pressure, and composition the range of possibilities is very large. Furthermore not all clouds condense directly from the gas to a solid or liquid phase as the same species. For example in the atmosphere of a gas giant exoplanet solid MnS cloud particles are expected to form around 1400 K from the net reaction H\textsubscript{2}S + Mn$\rightarrow $MnS(s) + H\textsubscript{2} (\textit{Visscher et al.} 2006).

Clouds strongly interact with incident and emitted radiative fluxes. The clouds of Earth and Venus increase the planetary Bond albedo (the fraction of all incident flux that is scattered back to space) and consequently decrease the equilibrium temperature. Clouds can also ``trap'' infrared radiation and heat the atmosphere. Hazes, in contrast, because of their usually smaller particle sizes can scatter incident light away from a planet but not strongly affect emergent thermal radiation, and thus predominantly result in a net cooling. The hazes of Titan play such an ``anti-greenhouse effect'' role in the energy balance of the atmosphere. For these reasons global atmospheric models of exoplanets, including those aiming to define the habitable zone given various assumptions, must consider the effects of clouds. However clouds are just one ingredient in such planetary atmosphere models. Bulk atmospheric composition, incident flux, gravity, chemistry, molecular and atomic opacities, and more must all be integrated along with the effect of clouds in order to construct realistic models. Introductory reviews by \textit{Burrows \& Orton} (2010) and \textit{Seager \& Deming} (2010) cover the important fundamentals atmospheric modeling and place cloud models in their broader context. 

In the remainder of this chapter we discuss the importance of clouds to exoplanet atmospheres, particularly considering their impact on habitability, discuss cloud modeling in general and the types of models developed for exoplanet studies, and finally briefly review observations of exoplanet clouds. Because clouds have played such a large role in efforts to understand the atmospheres of brown dwarfs, we also briefly review the findings of this field. We conclude with an overview of how the Mie opacity of particles behaves in various limits and consider the case of fluffy particles.

\section{Importance of Clouds and Hazes in Exoplanet Atmospheres}

\subsection{Albedo, Detectability, and Characterization}

Before discussing the role on clouds on the spectra of extrasolar planets it is worthwhile to review the various albedos which enter the discussion. The Bond or bolometric albedo is the fraction of all incident light, integrated over the entire stellar spectrum, which is scattered back to space by a planet. This albedo is a single number and enters into the computation of a planet's equilibrium temperature (the temperature an airless planet would have if its thermal emission were equal to the incident radiation which it absorbs). It is also useful to know the monochromatic ratio of all scattered to incident light as a function of wavelength, which is the spherical albedo. For historical reasons it is more common to discuss the geometric albedo, which is specifically the wavelength-dependent ratio of the light scattered by the entire planet in the direction directly back towards its star compared to that which would be so scattered by a perfectly reflecting Lambert disk of the same radius as the planet. Care must be taken to distinguish all these albedos as they can differ markedly from one another even for the same planet and the literature is rife with confusion between them.

Perhaps the single most important effect of clouds is to brighten the reflected light spectra of exoplanets, particularly at optical wavelengths ( $0.38<\lambda <1$ \textgreek{m}m). For planets with appreciable atmospheres, Rayleigh scattering is most efficient in the blue. At longer wavelengths however absorption by either the planetary solid surface or oceans (for terrestrial planets) or by atmospheric gaseous absorbers becomes important in the red. This is because for common molecules at planetary temperatures, such as H\textsubscript{2}O, CO\textsubscript{2}, or CH\textsubscript{4}, vibrational-rotational transitions become important at wavenumbers below about 15,000 cm\textsuperscript{{}-1} or wavelength  $\lambda >0.6$ \textgreek{m}m. Except for diatomic species (notably O\textsubscript{2}) planetary atmospheres are generally not warm enough to exhibit strong electronic absorption features in the optical. Thus the reflected light or geometric albedo spectrum of a generic cloudless planet with an atmosphere would be bright at blue wavelengths from Rayleigh scattering and dark in the red and at longer wavelengths from gaseous molecular or surface absorption.

Clouds, however, tend to be bright with a fairly gray opacity through the optical. Thus a thick, scattering cloud can brighten a planet in the far red by scattering more light back to space than a cloudless planet. As a result two similar planets, one with and one without cloud cover, will have very different geometric albedos in the red, and consequently differing brightness contrasts with their parent stars. The mean Earth water clouds increase the contrast between the reflection spectrum of an Earth-like planet and its host star by one order of magnitude (\textit{Kitzmann et al.} 2011b). This effect, first noted for giant planets by \textit{Marley et al. }(1999) and \textit{Sudarsky et al.} (2000) and further explored in \textit{Cahoy et al. }(2010) is illustrated in Figure 1 which plots geometric albedo as a function of  $\lambda $ for giant planets with and without clouds. A warm, cloudless atmosphere is dark in reflected light while a cooler atmosphere, sporting water clouds, is much brighter. In this case clouds affect the model spectra far more than a factor of three difference in the atmospheric abundance of heavy elements, which is also shown.

In the case of searches for planets by direct coronagraphic imaging in reflected light, clouds may even control whether or not a planet is discovered. Depending on the spectral bandpass used for planet discovery at a given distance from its primary star a cloudy planet may be brighter and more detectable than a cloudless one. Discussions of the influence of clouds and albedo on detectability include those by \textit{Tinetti et al. }(2006a) and \textit{Kitzmann et al. }(2011a).

\begin{figure}[h!]
  \centering
  \includegraphics[scale=0.5]{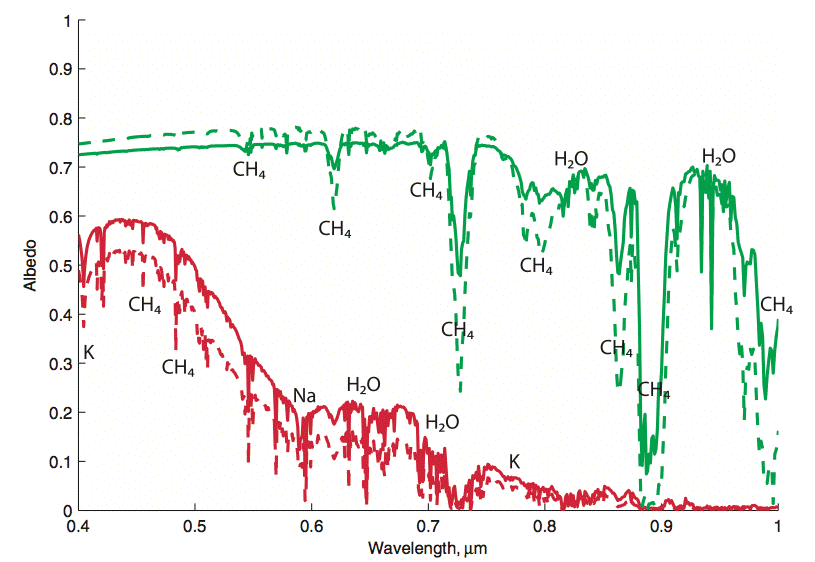} 
  \caption{Model geometric albedo spectra for Jupiter-like planets at 2 AU (green) and 0.8 AU (red) from their parent star. Solid and dashed lines lines show models with a solar abundance and a 3-times enhanced abundance of heavy elements respectively. Prominent absorption features are labeled. The 2 AU model planets possess a thick water cloud while the 0.8 AU models are warmer and cloudless and consequently darker in reflected light. Figure modified from \textit{Cahoy et al. }(2010).}
\end{figure}

Once a planet is detected, spectra are needed to characterize atmospheric abundances of important molecules. For terrestrial extrasolar planets this is best done by the analysis of the thermal emission spectrum (\textit{Selsis,} 2004; \textit{Tinetti et al.}, 2012). Clouds, however, may conceal the thermal emission from the surface and dampen spectral features of molecules (e.g., the bio-indicators N\textsubscript{2}O or O\textsubscript{3}). Indeed, clouds on Earth have a larger impact on the emitted infrared flux than the differences between night and day (\textit{Hearty et al.}, 2009; \textit{Tinetti et al.}, 2006a). Thermal emission spectra are therefore very sensitive to the types and fractional coverages of clouds present in the atmosphere. At high spectral resolution the most important terrestrial molecular spectral features in the mid-infrared, such as O\textsubscript{3}, remain detectable even for cloud covered conditions in many cases. Figure 2 shows thermal emission spectra affected by low-level water droplet and high-level water ice clouds of an Earth-like planet orbiting different main-sequence dwarf stars (adopted from \textit{Vasquez et al.}, 2013). With increasing cloud cover of either cloud type the important 9.6 \textgreek{m}m absorption band of ozone is strongly dampened, along with an overall decrease in the thermal radiation flux. For some cases presented in Fig. 2 (e.g. for the F-type star and 100\% high-level cloud cover) the ozone band seems to be completely absent or even appears in emission rather than absorption (F-type star, 100\% low-level clouds). At lower spectral resolution, such as could be obtained for terrestrial exoplanets in the near future, clouds render the molecular features even less detectable; thus clouds will strongly affect the determination of their atmospheric composition. For example, as shown by \textit{Kitzmann et al.} (2011a) a substantial amount of water clouds in an Earth-like atmosphere can completely hide the spectral signature of the bio-indicator ozone in low-resolution thermal emission spectra.

\begin{figure}[h!]
  \centering
  \includegraphics[scale=1.0]{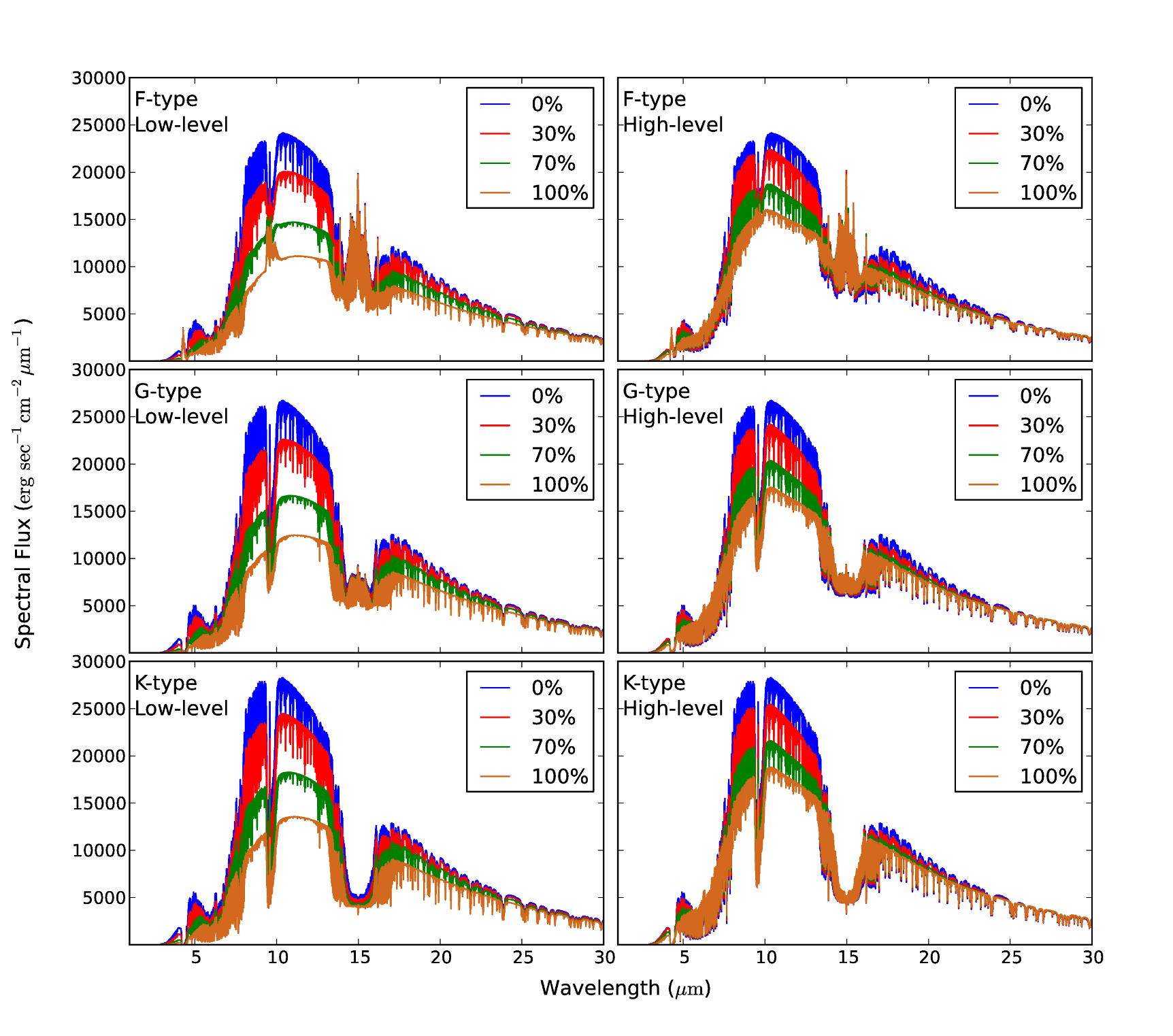} 
   \caption{Planetary thermal emission spectra influenced by low-level water droplet (left panel) and high-level ice clouds (right panel) for an Earth-like planet orbiting different kinds of main-sequence host stars (adopted from \textit{Vasquez et al. }2013). For each central star the spectra are shown for different cloud coverages. Note especially the strong impact of the cloud layers on the 9.6 \textgreek{m}m absorption band of ozone.}
	\end{figure}

Clouds can also obscure spectral features originating from the surface of a planet. In principle, signals of surface vegetation (``vegetation red edge'') are present in the reflected light spectra of Earth-like planets. However, as often pointed out (e.g., \textit{Arnold et al. }(2002) and \textit{Hamdani et al.} (2006)) this spectral feature can easily be concealed by clouds. For the detectability of possible vegetation signatures of terrestrial extrasolar planets \textit{Monta\~n\'es-Rodr\'iguez et al.} (2006) and \textit{Tinetti et al. }(2006b) concluded that clouds play a crucial role for these signatures in the reflection spectra. Thus, apart from the scattering characteristics of different planetary surface types, the presence of clouds has been found to be one of the most important factors determining reflection spectra. 

Transmission spectra of transiting planets can be used to obtain many atmospheric properties, such as atmospheric composition and temperature profiles. Transmission spectroscopy of extrasolar giant planets has already proven to be a successful method for the characterization of giant exoplanets. As shown by \textit{Pall\'e et al. }(2009) the major atmospheric constituents of a terrestrial planet remain detectable in transmission spectra even at very low signal-to-noise ratios. Thus, transmission spectra can provide more information about the atmospheres of exoplanets than reflection spectra. Theoretical transmission spectra of Earth-sized transiting planets have been studied by \textit{Ehrenreich et al. }(2006) including the effects of optically thick cloud layers. Their results show that the transmission spectra only contain information about the atmosphere above the cloud layer and that clouds can effectively increase the apparent radius of the planet. The impact of clouds on the transmission spectra therefore depends strongly on their atmospheric height. If cloud layers are only located in the lower atmosphere, which is already opaque due to absorption and scattering by gas species, their overall effect on the spectrum will be small (\textit{Kaltenegger \& Traub}, 2009).

\subsection{Habitability of Extrasolar Terrestrial Planets}

From the terrestrial bodies with an atmosphere in the solar system we know that clouds are a common phenomenon and should also be expected to occur in atmospheres of terrestrial extrasolar planets. Apart from the usual well-known greenhouse gases, clouds have the most important climatic impact in the atmospheres of terrestrial planets by affecting the energy budget in several ways. Firstly, clouds can scatter incident stellar radiation back to space resulting in atmospheric cooling (albedo effect). On the other hand, clouds can trap thermal radiation within the lower atmosphere by either absorption and re-emission at their local temperature (classical greenhouse effect) or by scattering thermal radiation back towards the surface (scattering greenhouse effect), which heats the lower atmosphere and planetary surface. All these effects are determined by the wavelength-dependent optical properties of the cloud particles (absorption and scattering cross-sections, single scattering albedo, asymmetry parameter, scattering phase function). These properties can differ considerably for different cloud forming species (owing to their refractive indices) and atmospheric conditions (composition and temperature structure). Note that the single scattering albedo, yet a fourth type of albedo (see Section 2.1 for the others), measures the fraction of all incident light scattered by a single cloud particle.

Life as we know it requires the presence of liquid water to form and survive. In the context of terrestrial exoplanets we are primarily concerned with habitable conditions on the planetary surface. Lifeforms may of course also exist in other environments, such as deep under the planetary surface or within a subsurface ocean (\textit{Lammer et al.}, 2009). Since there is no possibility of detecting the presence of such habitats by remote observations our current definition of a habitable terrestrial planet assumes a reservoir of liquid water somewhere on its surface. Liquid surface water implies that these planets would also have abundant water vapor in their atmospheres. We can therefore safely assume that water (and water ice) clouds will naturally occur in the atmospheres of these exoplanets throughout the classical habitable zone.

As such, H\textsubscript{2}O and CO\textsubscript{2} clouds are the prime focus for habitable exoplanets. Other possible condensing species also found in our solar system include C\textsubscript{2}H\textsubscript{6} and CH\textsubscript{4} (e.g., in the atmosphere of Titan), N\textsubscript{2} (Triton), or H\textsubscript{2}SO\textsubscript{4} (Venus, Earth). Any chemical species in liquid form on a planetary surface can in principle be considered as a potential cloud-forming species in the atmosphere if its atmospheric partial pressure is high enough and the temperature low enough.

Given these considerations it is clear that clouds are important for the determination of the boundaries of the classical habitable zone around different kinds of stars. In the next two subsections we discuss some of the ways that clouds influence the habitable zone boundaries.

\subsubsection{Inner Edge of the Habitable Zone}

The closer a terrestrial planet orbits its host star the greater stellar insolation and consequently surface temperature. This leads to enhanced evaporation of surface water and increases the amount of water vapor in the atmosphere. The strong greenhouse effect by this additional water vapor further increases the surface temperature and therefore the evaporation of surface water, resulting in a positive feedback cycle (see \textit{Covey et al.}, this volume).

The inner boundary of the classical habitable zone is usually defined either as the distance from a host star at which an Earth-like planet completely loses its liquid surface water by evaporation (runaway greenhouse limit), or as the distance from the host star when water vapor can reach the upper atmosphere (water loss limit). The latter is the definition favored by \textit{Kasting et al.} (1993); the former definition is used by \textit{Hart} (1979). In the present Earth atmosphere water escape is limited by the cold trap at the tropopause which loses its efficiency in a sufficiently warm, moist atmosphere (\textit{Kasting et al.}, 1993).

Using a one-dimensional atmospheric model, \textit{Kasting et al. }(1993) estimated that the inner boundary of the habitable zone for the runaway greenhouse scenario for an Earth around the Sun would be located at 0.84 AU. This distance, however, was calculated with the clouds treated as a surface albedo effect and neglecting any feedbacks. For example once formed clouds increase the planetary albedo which in turn partially reduces the temperature increase due to the higher stellar insolation and the cloud greenhouse effect. \textit{Kasting} (1988) investigated some feedback effects and concluded that with a single layer of thick water clouds Earth could be moved as close as 0.5 AU (for a cloud coverage of 100\%) or 0.67 AU (50\% cloud coverage) from the Sun before all liquid surface water would be lost. \textit{Goldblatt \& Zahnle }(2011) explored the various feedback issues in detail and concluded that more sophisticated modeling approaches are necessary to explore habitability.

Water clouds also contribute to the greenhouse effect. Depending on the temperature at which the cloud emits thermal radiation, its greenhouse effect can match or exceed the albedo effect. Cold clouds, such as cirrus, are net greenhouse warmers. Low clouds with temperatures near the effective radiating temperature affect climate almost entirely by their albedo, which depends on such factors as patchiness, cloud thickness, and the size distributions and composition of cloud particles. The exact climatic impact of clouds therefore crucially depends on the balance between their greenhouse and albedo effects. Whether the greenhouse or albedo effect dominates for clouds forming under runaway greenhouse conditions cannot be easily determined a priori. A better understanding of the cloud microphysics and convection processes in moist atmospheres during a runaway greenhouse process is needed to determine the cloud properties and their fractional coverage near the inner habitable zone (HZ) boundary. The effect of clouds under runaway greenhouse conditions represents one of the most important unresolved issues in planetary climate.

Note that this discussion is relevant to wet planets with moist atmospheres fed by extensive seas. Dry, land or desert, planets will have unsaturated atmospheres in the tropics and thus radiate at a higher temperature and cool more efficiently than planets with a water saturated atmosphere. As a result the habitable zone for such planets may be larger (\textit{Abe et al., }2011).

\subsubsection{Outer Edge of the Habitable Zone}

The outer edge of the habitable zone is set by the point at which there is no longer liquid water available at the surface as it is locked in ice. With falling insolation planets found progressively farther away from their central star have lower atmospheric and surface temperatures. With lower surface temperatures, the removal of CO\textsubscript{2} from the atmosphere due to the carbonate-silicate cycle, which controls the amount of CO\textsubscript{2} in Earth's atmosphere on time scales of order a million years, becomes less efficient (see \textit{Covey et al.}, this volume). Thus, if the terrestrial planet is still geologically active, CO\textsubscript{2} can accumulate in the atmosphere by volcanic outgassing. With decreasing atmospheric temperatures, CO\textsubscript{2} will condense at some point to form clouds composed of CO\textsubscript{2} ice crystals. Because condensation nuclei can be expected to be available in atmospheres of terrestrial planets, the most dominant nucleation process for the formation of CO\textsubscript{2} clouds is usually assumed to be heterogeneous nucleation (\textit{Glandorf et al 2002}), in which cloud particles form on existing seed particles.

Just like water clouds, the presence of CO\textsubscript{2} clouds will result in an increase of the planetary albedo by scattering incident stellar radiation back to space. However, in contrast to water, CO\textsubscript{2} ice is almost transparent in the infrared (\textit{Hansen}, 1997, 2005) except within some strong absorption bands. Thus CO\textsubscript{2} clouds are unlikely to trigger a substantial classical greenhouse effect by absorption and re-emission of thermal radiation.

On the other hand, CO\textsubscript{2} ice particles can efficiently scatter thermal radiation. This allows for a scattering greenhouse effect in which a fraction of the outgoing thermal radiation is scattered back towards the planetary surface (\textit{Forget \& Pierrehumbert}, 1997). Depending on the cloud properties this scattering greenhouse effect can outweigh the albedo effect and can in principle increase the surface temperature above the freezing point of water. However, the scattering greenhouse effect is a complex function of the cloud optical depth and particle size (\textit{Colaprete and Toon, }2003). Furthermore, because the greenhouse effect of CO\textsubscript{2} clouds depends on the scattering properties of the ice particles, the particle shape (which cannot be expected as spherical) or particle surface roughness may play an important role. However, these effects cannot be easily quantified because neither the particle shapes nor their surface properties are known. These and other uncertainties in the cloud microphysics of CO\textsubscript{2} ice in cool CO\textsubscript{2} dominated atmospheres makes the calculation of the position of the outer HZ boundary complicated. More details on the formation of CO\textsubscript{2} clouds in the present and early Martian atmosphere can be found in \textit{Glandorf et al. }(2002), \textit{Colaprete \& Toon }(2003), and \textit{M\"a\"att\"anen et al. }(2005), as well as in \textit{Esposito et al.} (this volume). Inferences about CO\textsubscript{2} cloud particle sizes in the current martian atmosphere as constrained by a variety of datasets are presented by \textit{Hu et al.} (2012).

For a fully cloud-covered early Mars with thick a CO\textsubscript{2} dominated atmosphere and CO\textsubscript{2} clouds composed of spherical ice particles \textit{Forget \& Pierrehumbert }(1997) estimated that the outer boundary of the HZ is located at 2.4 AU, in contrast to the cloud-free boundary of 1.67 AU by \textit{Kasting et al. }(1993). This greater value has been further used by \textit{Selsis et al}. (2007) to extrapolate the effects of CO\textsubscript{2} clouds on the outer HZ boundary towards other main sequence central stars (see also \textit{Kaltenegger \& Sasselov}, 2011).

For terrestrial super-Earths it has been suggested by \textit{Pierrehumbert \& Gaidos }(2011) that these planets could have retained much of their primordial H\textsubscript{2} atmosphere owing to their greater mass. According to their model study, the classical habitable zone might be far larger than expected for a CO\textsubscript{2} dominated atmosphere although this study did not explore the impact of clouds.

\section{Clouds and Radiation}

Clouds are important for planetary atmospheres because they interact with both incident short wavelength radiation from the parent star and emergent thermal radiation emitted by the planetary surface, if present, and the atmosphere itself. Perhaps the simplest example of such interaction occurs for spherical cloud particles composed of a single constituent. In this case the wavelength dependent optical properties can be computed from Mie calculations that depend only on particle size and the wavelength-dependent complex refractive index of the bulk condensate. In this section we briefly review the basics of the interaction of cloud particles with radiation and summarize the use of Mie theory for modeling this interaction as well as point out some useful simplifications that can be applied in certain limiting cases.

\subsection{Basic Radiative Properties of Cloud Particles}

Cloud opacity ultimately depends upon the radiative properties of the constituent particles. A particle with radius \textit{r} has a cross section to scatter losslessly or to absorb incident radiation at some wavelength \textit{\textgreek{l}}, given by \textit{C}\textit{\textsubscript{s}} or \textit{C}\textit{\textsubscript{a}} respectively. These cross sections are defined as  $C_s=Q_s\pi r^2$ and $C_a=Q_a\pi r^2$, and their sum is the extinction cross section \textit{C}\textit{\textsubscript{e}}. The scattering and absorption efficiencies \textit{Q}\textit{\textsubscript{s}} and \textit{Q}\textit{\textsubscript{a}}, and, from them, the extinction efficiency \textit{Q}\textit{\textsubscript{e}} = \textit{Q}\textit{\textsubscript{s}}\textit{ }+ \textit{Q}\textit{\textsubscript{a}} is thus defined. The particle single scattering albedo is $\omega =Q_s/Q_e$, and the scattering phase function is  $P\left(\theta \right)$. These quantities can be computed by a Mie code, which computes the electromagnetic wave propagation through spherical particles, given a tabulation of the optical properties of the cloud material. Good physically-based introductions to radiative properties of particles can be found in \textit{van de Hulst }(1957), \textit{Hansen \& Travis }(1974), and \textit{Liou} (2002).

Modeling cloud particle radiative properties can seem forbidding; workers often think a Mie scattering code, with all its exotic predictions, is needed. However, in many applications, much simpler approaches not only can provide very good quantitative fidelity - and greater physical insight - but also can be applied directly to more complex kinds of particles than the spheres for which Mie theory is derived.

The direct beam of radiant flux vertically traversing a cloud \textit{layer} of particles composed of some material \textit{q }is reduced by a factor $\exp \left(-\tau _q\right)$, where the layer optical depth is  $\tau _q=nQ_e\pi r^2H_q=nC_eH_q$, and \textit{n} is the particle number density, \textit{r} the particle radius, and \textit{H}\textit{\textsubscript{q}} the vertical thickness of the layer (assuming a uniform vertical distribution for simplicity). \textit{Q}\textit{\textsubscript{e}}\textit{ }and thus  $\tau _q$ are functions of the wavelength \textit{\textgreek{l}}, through the \textit{\textgreek{l}}{}--dependent real and imaginary refractive indices of the material in question (\textit{n}\textit{\textsubscript{r}}\textit{ , n}\textit{\textsubscript{i}}\textit{;} see below, and \textit{Draine \& Lee, }1984 or\textit{ Pollack et al., }1994 or \href{http://www.astro.uni-jena.de/Laboratory/Database/databases.html}{\textstyleInternetlink{ }}\url{http://www.astro.uni-jena.de/Laboratory/Database/databases.html} for typical values). The particle opacity \textit{\textgreek{k}} (in units of length-squared per mass) is the effective particle cross section per unit mass of either solids or gas.

If the phase function is strongly forward scattering, as can be the case for particles with  $r/\lambda >10$ or so (\textit{Hansen \& Travis }1974), much of the ``scattering'' can be approximated as unattenuated radiation by reducing the layer optical depth and renormalizing the particle albedo and phase function. This is done by applying simple corrections to both \textit{\textgreek{w}} and  $\tau_q$ known as similarity relations. \textit{Irvine, (}1975), \textit{van de Hulst} (1980), and \textit{Liou} (2002) review these, and present a number of ways to solve the radiative transfer problem in general.

\subsection{Refractive Indices and Opacity}

All calculations of particle radiative properties (\textit{Q}\textit{\textsubscript{e}}, \textit{Q}\textit{\textsubscript{s}}, \textit{Q}\textit{\textsubscript{a}},  $\omega $ and \textit{P(\textgreek{j)}}) begin with the refractive indices of the material and the particle size. While Mie codes are available for download (e.g\textit{.}, ftp://climate1.gsfc.nasa.gov/wiscombe/Single\_Scatt/), analytical approximations that are valid in all the relevant limits can often be of great use. Such approximations rely on the fact that realistic clouds have size and shape distributions that average away the exotic fluctuations shown by Mie theory for particles of specific sizes (see \textit{Hansen \& Travis }1974 for examples). With current computational capabilities, Mie calculations are not onerous, but if many wavelengths and/or grids of numerous models are of interest, the burden is compounded, so one needs to understand whether such detailed predictions are actually needed.

One simple approximation to scattering by equidimensional, but still irregular, particles was developed by \textit{Pollack \& Cuzzi} (1980). This approach uses Mie theory for particles with small-to-moderate $r/\lambda$, where shape irregularities are indiscernible to the waves involved and Mie calculations converge rapidly. For larger $r/\lambda $ a\textit{ }combination of diffraction, external reflection, and internal transmission is used as adjusted for shape and parametrized by laboratory experiments. The approach is easy to apply in cases where the angular distribution of scattered radiation is important, as long as a particle size distribution smears away the significant oscillations in scattering properties which Mie theory predicts for monodisperse spheres (\textit{Hansen \& Travis }1974). This method saves on computational effort for particles with large $r/\lambda $ but does involve a Mie code for the smaller particles.

Even simpler approaches are feasible. In many cases of interest, only globally averaged emergent fluxes or reflectivities, or perhaps heating calculations, are needed; here, the details of the phase function are of less interest than the optical depth and particle albedo, which are based only on the efficiencies. For these cases one can do fairly well using asymptotic expressions for efficiencies in the limiting regimes  $r/\lambda {\ll}1$ and  $r/\lambda {\gg}1$, connecting them with a suitable bridging function. These expressions depend on the refractive indices of the material in question.

The simplest limit is when  $r/\lambda {\gg}1$ (geometrical optics); in fact, many of the cloud models discussed in this chapter fall into this regime at 1-10 \textgreek{m}m wavelengths. In this range it is convenient to neglect diffraction, which is strongly concentrated in the forward direction and can be lumped with the direct beam as noted above. Then, assuming the particle has density \textgreek{r }and is itself opaque, $Q_e{\leq}1$ and \textit{C}\textit{\textsubscript{e}}\textit{ }reduces to the physical cross section. Thus the solids-based opacity is simply  $\kappa =3\pi r^2/4\rho \pi r^3=3/4r\rho $. In this regime, particle growth drastically reduces the opacity because the surface to mass ratio decreases linearly with radius (\textit{Miyake et al.,} 1993; \textit{Pollack et al., }1994).

At the opposite extreme, when  $r/\lambda {\ll}1$(the Rayleigh limit), simple analytical expressions also apply (\textit{Draine \& Lee }1984; \textit{Bohren \& Huffman }1983). For simplicity below we give the expressions for  $n_i{\ll}n_r{\sim}1$ (appropriate for silicates, oxides, water, but not iron), but the general expressions are not much more complicated (\textit{Draine \& Lee }1984). Thus  $Q_a=24xn_rn_i/\left(n_r^2+2\right)^2$ and  $Q_s=8x^4\left(n_r^2-1\right)^2/3\left(n_r^2+2\right)^2$, where  $x=2\pi r/\lambda $(see also \textit{van de Hulst, }1957, p. 70). Because \textit{Q}\textit{\textsubscript{s}} decreases much faster than\textit{ Q}\textit{\textsubscript{a}} with decreasing  $r/\lambda $ in this regime\textit{, }small\textit{ }particles become not only less effective at interacting with radiation, but increasingly absorbers/emitters rather than scatterers. In this limit, their cross section  $C_a=Q_a\pi r^2$ per unit particle mass becomes constant since \textit{Q}\textit{\textsubscript{a}}\textit{ }is proportional to \textit{r}.

\subsection{Shape and Porosity}

In the case of terrestrial cloud droplets, in which condensation dominates growth, and raindrops, in which coagulation dominates growth, spherical particles of constant density are assumed, provided they fall slowly (though drag-induced deformation is taken into account when computing terminal fall-speeds for large raindrops). However, if particles condense from their vapor phase as solids, tiny initial monomers may instead coagulate by sticking into porous aggregates, perhaps having some fractal properties where the density may depend on the size. Non-spherical particle shape adds complexity to the computation of optical properties, which is commonly the case for solid particles. For instance, the single-scattering properties of ice particles in the terrestrial atmosphere depend on not only the geometric shape of the crystals (which depend on the temperature), but also the microscopic surface roughness. As noted by \textit{Fu et al. }(1998), the extinction and absorption cross sections depend mainly on projected areas and particle volumes (note that random orientation is typically assumed), while the scattering phase function are largely determined by the aspect ratio of the crystal components and their small-scale roughness (\textit{Fu}, 2007).

In the general exoplanet case for tiny particles in gas of typical densities, plausible inter-particle collision velocities are very small and lead to sticking but only minor compaction (see e.g., \textit{Cuzzi and Hogan} 2003, \textit{Dominik and Tielens} 1997) until particles become large enough (10-100 \textgreek{m}m) where sedimentation velocity can exceed several meters per second and compaction or bouncing arise (section 6.3; see \textit{Dominik et al. 2007}, \textit{Zsom et al.} 2010). The thermodynamic properties of condensates and the temperature-pressure (\textit{T-P) }structure of the atmosphere in question determine whether the condensate appears as a liquid or a solid. Figure 3\textbf{ }shows condensation curves of a number of important cloud-forming compounds, along with the \textit{T-P} profiles of a range of exoplanetary and substellar objects.

Models of growing aggregates have been fairly well developed in the literature of protoplanetary disks (e.g.\textit{, Weidenschilling,} 1988; \textit{Dominik \& Tielens, }1997; \textit{Beckwith et al., }2000; \textit{Dominik et al., }2007; \textit{Blum,} 2010). The properties of clouds made of this rather different kind of particle will differ in their opacity and vertical distribution from predictions of the simplest models; these properties in turn will impact, and can in principle be constrained by, remote observations of reflected and emitted radiation.

\begin{figure}[h!]
  \centering
  \includegraphics[scale=0.8]{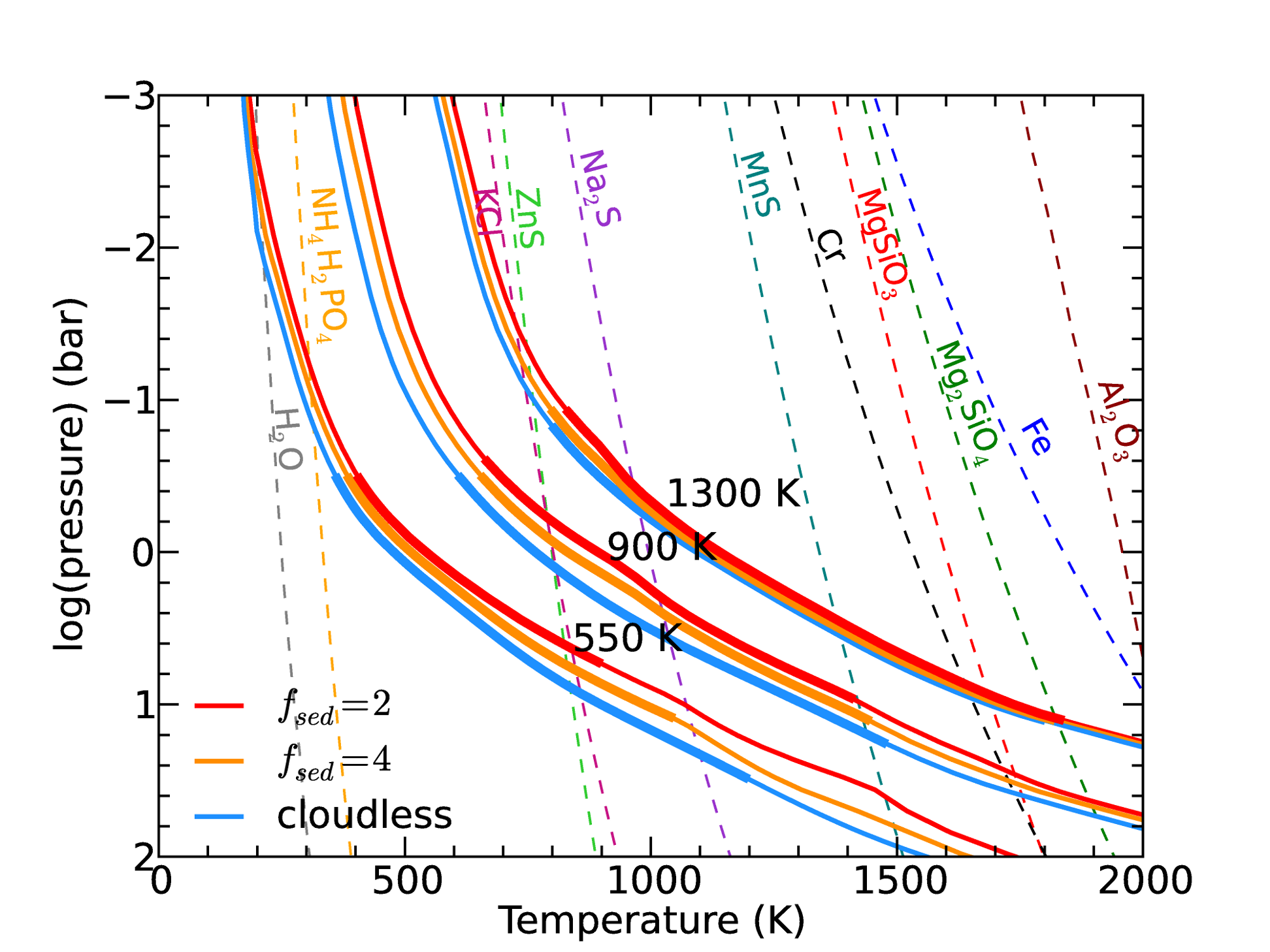}
  \caption{Model atmospheric temperature-pressure profiles for brown dwarfs (solid) and condensation curves (dashed) for a variety of compounds. For each model effective temperature three curves are shown, corresponding to a cloudless calculation (blue) and two cloudy models with varying sedimentation efficiency, \textit{f}\textsubscript{sed }= 2 (red) and 4 (orange). Figure modified from \textit{Morley et al. }(2012).}
\end{figure}

\subsection{Porous Particles and Effective Medium Theory }

The most straightforward way of modeling porous aggregates is to model their effective refractive indices based on their constituent materials and porosity. If the monomers from which the particles are made are smaller than the wavelength in question, they act as independent dipoles immersed in an enveloping medium (the medium can be another material; here we assume it is vacuum). The aggregate as a whole can then be modeled as having \textit{effective} refractive indices which depend only on the porosity of the aggregate and the refractive indices (but not the size) of the monomers. This is the so-called \textit{effective medium theory (EMT); }several variants are discussed by \textit{Bohren \& Huffman }(1983), \textit{Ossenkopf }(1991), \textit{Stognienko et al}. (1995), and \textit{Voschinnikov et al.} (2006).

For most combinations of materials, EMTs can be even further approximated by simple volume averages such that the refractive indices of the particle as a whole can be written (for a simple two-component particle with component 2 being vacuum) as  $n_i=fn_{\mathit{i1}}+\left(1-f\right)n_{\mathit{i2}}$ and $\left(n_r-1\right)=f\left(n_{\mathit{r1}}-1\right)+\left(1-f\right)\left(n_{\mathit{r2}}-1\right)$, where \textit{n}\textit{\textsubscript{r }}and\textit{ n}\textit{\textsubscript{i}} are the effective real and imaginary indices of the aggregate as a whole, and component 1 has volume fraction \textit{f}. If a material which has large refractive indices, such as iron, is involved, the full expressions are needed (see \textit{Wright}, 1987; \textit{Bohren \& Huffman}, 1983; or other basic references). We will assume the porous aggregates are roughly equidimensional, not needle-like structures, but even that added complication can often be tractable in semi-analytical ways for most materials (\textit{Wright} 1987; \textit{Bohren \& Huffman }1983).

In the $r/\lambda {\ll}1$ regime, porosity plays little role, both because the opacity is simply proportional to the total mass regardless of how it is distributed (see above), and because ever-smaller particles are unlikely to be aggregates of ever-smaller monomer constituents, but are more likely to be monomers themselves. \textit{Ferguson et al.} (2007) present numerical calculations which show that, for the tiny particles they modeled (0.35 nm - 0.17 \textgreek{m}m) indeed this expectation is fulfilled to first order, also showing that mixing of materials within these tiny aggregates has little effect (see also \textit{Allard et al.} 2001, \textit{Helling et al.} 2008, \textit{Witte et al.} 2009 and references therein). Particles this small are likely to be well-mixed at all levels of typical exoplanet atmospheres (see below). Some targets of interest may have high haze layers where this regime is of interest.

However, in the  $r/\lambda {\gg}1$ regime, which covers typical planetary clouds observed in the optical or near-infrared, porosity does affect opacity. Specifically, for a particle of mass \textit{M} and porosity $\varphi$, the opacity  $\kappa =C_e/M=3/4\rho r=3/4\rho _s\left(1-\varphi \right)r$ where\textit{ } $r=\left(3M/4\rho _s\left(1-\varphi \right)\right)^{1/3}$\textit{ }is the actual radius. Comparing the opacity of this particle to that of a ``solid'' particle with the same mass having radius\textit{ r}\textit{\textsubscript{s}}\textit{ , }gives $\kappa /\kappa _s=\left(1-\varphi \right)^{-2/3}$. The porous particle thus has a larger opacity for the same mass, and indeed reaches the geometrical optics (\textit{\textgreek{l }}independent) limit at lower mass than a nonporous particle. In this regime, effective medium theory should be a convenient and valid way of modeling porous and/or aggregate grain properties (section 3.2; see also \textit{Helling et al.} 2008 and references therein).

Variable composition introduces further complexity to computing single-scattering properties, requiring a mixing rule to compute the refractive index for the composite particle. Different mixing rules can introduce uncertainties of order 1 and 10\% respectively in the real and imaginary refractive indices for black carbon inclusions in liquid water droplets (\textit{Lesins et al}., 2002).

\section{Cloud Models}

The cloud properties required to compute radiative fluxes are rather extensive. Thermal emission requires the emissivity and temperature of the cloud particles. Vertical fluxes can be computed from knowledge of the vertical distribution of particle cross-section (for each species of particle) together with the wavelength dependent scattering phase function and the single-scattering albedo. For a basic two-stream approach these quantities can be boiled down to a vertical profile of extinction coefficient, asymmetry parameter, and single-scattering albedo.

Complexities can arise if particle temperature differs from the local atmospheric temperature by virtue of latent heat exchange and radiative heating of the particles, but this should be rare in exoplanet applications (\textit{Woitke \& Helling }2003) and typically the particles are assumed to be at ambient atmospheric air temperature. Clouds that are not horizontally uniform are another possible complication that can be considered.

The required scattering and absorption total cross sections and overall asymmetry parameter needed for input to a radiative transfer model are found by integration of the single-scattering properties and emissivities over the particle size distributions. The size distributions in turn can be computed by a so-called bin model which tracks the number of particles in multiple different bin sizes as the particles interact with the atmosphere. Such an approach is computationally expensive, and a more efficient treatment is to assume a particular functional form with a small number of free parameters.

For terrestrial applications log-normal distributions, with three parameters (total number, geometric mean radius, and geometric standard deviation) and gamma distributions, also with three parameters (total number, slope parameter, and shape parameter) are commonly used. Separate size distributions are used for particles of different phases and bulk densities (e.g., raindrops and snowflakes) and for each mode of a multimodal size distribution. For example, the parameterized size distribution of cloud droplets, which grow principally through condensation in the terrestrial atmosphere, are treated distinctly from raindrops, which grow principally through collisions. Although a poor approximation for cloud droplets, it is often assumed that the shape parameter is zero for other cloud species, which allows the gamma distribution to collapse to an exponential distribution, thereby reducing the free parameters from three to two.

For the computation of heating rates from the divergence of radiative fluxes, the vertical distribution of any absorbing and emitting species is obviously critical. Also, any vertical redistribution by scatterers in bands with emission or absorption requires that their vertical distribution is accurate. If horizontal photon transport is unimportant, it is feasible to represent the global atmosphere-{}-or the atmosphere within any model column of finite horizontal extent, for that matter-{}-with two columns, one clear and one cloudy (for example see the discussion in the context of brown dwarfs in \textit{Marley et al.} (2010)). Assumptions about the vertical overlap of clouds are critical and can induce large errors in top-of-atmosphere fluxes as well as heating rates (e.g., \textit{Barker et al., }1999).

In the context of extrasolar planet atmosphere modeling we must connect a simple, usually 1D model of the atmosphere with what is potentially a complex brew of cloud properties which we would ideally need to know. However it is clear that the number of variables can quickly grow to unmanageable extent. In this chapter we discuss various approaches that have been taken to address this problem.

\subsection{Conceptual Framework for Cloud Modeling}

Given a profile of atmospheric temperature as a function of elevation or pressure we can ask where clouds form and what are their radiative properties. Here we give a simplified discussion of the problem following \textit{Sanchez-Lavega et al. }(2004).

The abundance of a given atmospheric species \textit{a} in the vapor can be given by its mass mixing ratio  $m_a=\rho _a/\rho =\epsilon P_a/P$ where $\rho $ and \textit{P} are the density and pressure of the {}``background'' atmosphere (that is, not including species \textit{a}) and  $\rho _a$ and \textit{P}\textit{\textsubscript{a}} are the density and partial pressure of vapor species \textit{a}. Here $\epsilon =\mu _a/\mu $ or the ratio of the molecular weight of species \textit{a, } $\mu _a$, to the background gas mean molecular weight  $\mu $. A not unreasonable assumption is that any vapor in excess of the saturation vapor pressure of \textit{a, } $P_{v,a}\left(T\right)$, condenses out. We can define the saturation ratio of \textit{a} as  $f_a=P_a/P_{v,a}\left(T\right)$.

Deep in the atmosphere, below cloud base,  $P_a{\ll}P_{v,a}$ so the species is found in the gas phase. In a Lagrangian framework one can imagine a rising parcel of gas that cools adiabatically as it expands, and at some point it may reach $P_a=P_{v,a}$ and a saturation ratio  $f_a=1$. In an Eulerian framework this condition requires that the thermal profile intersects with the vapor pressure curve for a given constituent, for instance as the 550 K model intersects the H\textsubscript{2}O vapor pressure curve in Figure 3. However the thermal profile for the 1300 K model does not intersect the H\textsubscript{2}O vapor pressure curve. Thus H\textsubscript{2}O would be expected to condense in the cooler, but not the warmer, atmosphere. To this point the problem is relatively straightforward for those species that condense directly from the gas phase, such as water. For other species, however, condensation instead proceeds by a net chemical reaction when the condensed phase has a lower Gibb's free energy than the vapor phase. One such example is H\textsubscript{2}S + 2Na $\rightleftharpoons $ Na\textsubscript{2}S(s) + H\textsubscript{2}.

Above the condensation level a number of issues arise. First, the assumption that all vapor in excess of saturation condenses may be faulty. Condensation may require a supersaturation $\left(f_a>1\right)$ before it proceeds, owing to the thermodynamic energy barrier of forming new particles. If so, what degree of supersaturation is required? Above cloud base, which is the lowest level at which condensation occurs, abundance of the condensed phase will depend on a balance between the convective mixing of particles from below and the downward sedimentation of the condensate particles. If the sedimentation rate of some portion of the size distribution of condensate is greater than the vertical mixing velocity scale, the scale height for the condensed phase, \textit{H}\textit{\textsubscript{a}}, will be smaller than the atmospheric scale height \textit{H} evaluated at cloud base. In the solar system typical values of \textit{H}\textit{\textsubscript{a}}\textit{/H} range from 0.05 to 0.20 (\textit{Sanchez-Lavega et al. }2004).

Nucleation is the starting process for the formation of a condensed phase (either liquid or solid) from a gaseous state, or the formation of solid from a liquid state. It creates an initial distribution of nuclei (embryos) which, if large enough, are stable with respect to the higher-entropy phase and tend to grow into larger particles by condensation or freezing. Such a phase transition can only occur spontaneously under thermodynamically favorable conditions (for the following discussion we first focus on the process of condensate nucleation from the vapor phase). Such conditions require the saturation ratio  $f_a$ to exceed unity ($f_a=1$ characterizes phase equilibrium).

The simplest nucleation mechanism is homogeneous nucleation, where the initial nuclei are formed by random spontaneous collisions of monomers in the vapor phase (e.g. single H\textsubscript{2}O\textsubscript{ }molecules) into a molecular cluster. This process is, however, connected to an energy barrier that can prevent the formation of a stable new phase. A supersaturated gas phase possesses a high Gibb's free energy, whereas at the same time molecules in the condensed phase would be at a lower potential energy. Thus, removing molecules from the vapour phase and adding them into a condensed phase would in principle lower the total free energy of the entire thermodynamical system. The corresponding net change in free energy depends mainly on the volume of an embryo (as a function of the particle radius  $r$) and the supersaturation of the gas.

However, the creation and growth of nuclei also implies that a new surface is generated which additionally contributes to the free energy owing to surface tension. It is only thermodynamically favourable to form a condensed phase when the net reduction of free energy in the system $\left({\sim}r^3\right)$ is greater than (or equal to) the free energy required from the surface tension $\left({\sim}r^2\right)$. Therefore, there exists a critical radius for which these two contributions balance each other. Embryos smaller than the critical size are unstable and will tend to evaporate again quickly whereas those with sizes larger than the critical radius will tend to grow freely. The critical radius is, in particular, a function of the saturation ratio. Low supersaturations yield very large critical radii while, on the other hand, for increasing values of $f_a$ the size of the critical embryos decreases.

Homogeneous nucleation usually requires a very high supersaturation and is therefore unlikely to occur in atmospheres of terrestrial planets on a large scale (observed supersaturations for water vapor in the atmosphere of Earth are normally in the range of a few percent, for example), because heterogeneous nucleation occurs at much lower supersaturations and thus quenches the supersaturation well before homogeneous nucleation occurs.

The predominant nucleation process from the vapor for terrestrial planetary atmospheres is thought to be heterogeneous nucleation. Here, the initial distribution of nuclei is formed on pre-existing surfaces. The presence of such surfaces substantially lowers the supersaturation required to form the critical clusters. Depending on the properties of these surfaces (e.g. whether they are soluble with respect to the condensed phase) nucleation is already possible for saturation ratios close to unity. Such surfaces can be provided by dust, sea salt, pollen, or even bacteria. In case of terrestrial planets these particles (condensation nuclei) are usually largely available because of mechanical process associated with the planetary surface-{}-such as wave breaking, bubble bursting, and dust saltation-{}-and from the formation of haze particles (such as sulfuric acid or sulfate droplets) that result from photochemical processes. Therefore, heterogeneous nucleation can reasonably be expected as the dominant nucleation mechanism for terrestrial extrasolar planets. This assumption, however, complicates the treatment of cloud formation because details on such condensation nuclei (composition, number density, size distribution) are not available in case of exoplanets.

Additionally, the formation of a solid phase (e.g. ice crystals) can in principle occur by different pathways. It can either form directly from the gas phase by homogeneous or heterogeneous nucleation. On the other hand, it is also possible to form the liquid phase as an intermediate step, followed by freezing of the supercooled liquid into the solid phase afterwards. Whether this indirect or the direct pathway occurs depends on the properties of the condensing species and on the local atmospheric conditions. Water ice cloud crystals in the Earth's atmosphere form by homogeneous and heterogeneous freezing of liquid in mixed-phase clouds (such as cumulonimbus and Arctic stratiform clouds) and heterogeneous nucleation in cirrus clouds. In the Martian atmosphere CO\textsubscript{2} clouds form directly by heterogeneous nucleation from the gas phase.

Given a composition of the cloud the task for any cloud model then becomes one of computing cloud particle sizes and their number distribution through the atmosphere above cloud base. A very simple solution would be to simply assume a mean particle size and a cloud scale height and this is effectively the approach many investigations have take to explore the effect of exoplanet clouds. For the remainder of this section we consider efforts to more rigorously derive expected particle sizes and the vertical distribution of cloud particles. We begin by considering the most thoroughly modeled clouds, terrestrial clouds of liquid water and water ice. We then move on to cloud models that have been constructed for terrestrial and giant extrasolar planets and conclude by considering the lessons learned from efforts to model clouds expected in brown dwarf atmospheres.

\subsection{Perspective from Earth Science}

Cloud modeling of terrestrial clouds comes in a wide assortment of classes. For detailed cloud studies of limited spatial extent, dynamical frameworks range from 0-D parcel models, to 1-D column models, to 2-D eddy-resolving models, to 3-D large-eddy simulations and cloud-resolving models. The difference between such models is the number of spatial dimensions represented in the governing equations. Another varying aspect of cloud models is the degree of detail in describing cloud microphysics. The simplest models simply assume that all vapor in excess of saturation condenses and assume a fixed size for the cloud particles. Others assume a functional shape for the cloud particle size distributions and parameterize the microphysical processes, prognosing one, two, or three moments of the size distribution for each condensate species (such as cloud water, rain, cloud ice, snow, graupel, and hail). The most detailed approach is to resolve the cloud particle size distributions without making any assumptions about the functional shape of the size distributions.

Global climate model frameworks similarly range from one-dimensional radiative convective models (e.g., \textit{Manabe \& Wetherald}, 1967), to two-dimensional, zonally averaged models (e.g., \textit{Schneider}, 1972), to modern three-dimensional general circulation models (GCMs). (See \textit{Schneider \& Dickinson}, 1974 for an early, comprehensive review of approaches to global climate modeling.) With respect to clouds in global models, one-dimensional radiative convective models suffer from a major deficiency, namely predicting plausible estimates of horizontal cloud coverage (see \textit{Ramanathan \& Coakley}, 1978). A two-dimensional framework is an intermediate step, though modern climate studies rely principally on three-dimensional GCMs. The rest of this section will focus on clouds in modern GCMs.

The representation of clouds in GCMs remains a major challenge, as cloud feedbacks constitute a leading source of uncertainty in current model-derived estimates of overall climate sensitivity, which are typically cast in terms of the sensitivity of globally averaged surface temperature to changes in radiative forcings (see \textit{Hansen et al.,} 1984). The response of tropical cirrus clouds to increasing sea surface temperatures has been a topic of great debate in the last two decades. At one extreme is the thermostat hypothesis of \textit{Ramanathan \& Collins} (1991), which suggests that increased water vapor leads to more extensive, thicker anvils that will on net cool the planet through increased albedo, while at the other extreme is the argument of \textit{Lindzen et al. }(2001) that greater condensate loading in a warmer climate precipitate more efficiently and lead to a dryer upper atmosphere that traps less infrared energy, driving a negative water vapor feedback. Both hypotheses produce a negative climate feedback for tropical clouds, one employing increased solar reflection, the other relying on increased infrared emission. While these are provocative ideas that have spawned countless evaluations of the hypothesis (with neither withstanding scrutiny), the predominant concept currently is that the tropical climate feedback for modern GCMs is determined primarily by the response of clouds in the marine boundary layer (\textit{Bony \& Dufresne,} 2005), of which the transition from overcast stratocumulus to broken cloud fields of trade cumulus is a leading primary candidate responsible for that response. With respect to cloud feedbacks generally, attention on the climate feedback of cirrus clouds formed from the detrainment of deep tropical convection has been supplanted to a large degree by more recent focus on the climate feedback of shallow clouds.

A fundamental problem in representing clouds in the GCMs is that the native GCM grid cells are very coarse, of order 100 km horizontally and 1 km vertically. Although model resolution steadily improves as computing power advances, the problem of convection and clouds being unresolved persists in models designed to span the globe and simulate climate over time scales of decades to millennia. The basic assumption in a conventional treatment of convection and clouds in a GCM is that cloud properties and precipitation rates in a model grid cell, which is much larger horizontally than any cloud, can be computed based on the mean properties of the grid cell. The purpose of a cloud parameterization is to compute cloud properties and precipitation rates from those mean properties.

A somewhat recent version of the NASA GISS (Goddard Institute for Space Studies) GCM (\textit{Schmidt et al.,} 2006) includes a conventional treatment of clouds and convection. The atmospheric model grid spacing is roughly 2 {\texttimes} 2.5{\textdegree} with 20 or 23 vertical layers, and thus all convection and cloud physics are necessarily highly parameterized. Deep convection is parameterized based on the convective instability of model columns using idealized updrafts and downdrafts, which detrain air into stratiform cloud layers. The stratiform cloud cover is computed as a diagnostic function of grid-scale relative humidity, and the relative humidity thresholds used to compute cloud cover in that diagnostic function serve as the principle tuning knobs for the GCM, with the dual tuning targets of top-of-atmosphere radiative balance and an overall albedo reasonably close to satellite-based estimates (note that such tuning can easily result in exaggerated cloudiness in some regions that make up for insufficient cloudiness in others). In the GISS GCM the only prognostic cloud variable is the mass mixing ratio of cloud condensate, which is a fundamental component of the cloud parameterization. Any precipitation is assumed to evaporate or fall out in one time step (30 minutes) and the phase of the condensate is probabilistically determined from temperature to allow for a modest amount of supercooling (liquid colder than the melting point) on average. The standard version of the GCM assumes different cloud droplet concentrations over ocean and land and also assumes a fixed number concentration for ice particles, and a fixed effective variance of the condensate size distributions is assumed for computing cloud optical thickness.

A more complex approach for stratiform cloud microphysics is used in the most recent version of the NCAR (National Center for Atmospheric Research) GCM. That scheme uses two moments (mass and number) for two prognostic (cloud water and cloud ice) and two diagnostic (rain and snow) hydrometeor species (\textit{Morrison \& Gettelman,} 2009). The rapidly sedimenting species are treated diagnostically with a tridiagonal solver to allow for long time steps (20 minutes with 2 microphysics substeps) in a manner that avoids numerical instability associated with falling through more than one layer during a time step. A novel aspect of this microphysics scheme is that by assuming a particular subgrid-scale distribution of cloud water, microphysical processes that involve cloud water take into account the problem of grid-averaging over nonlinear process rates (\textit{Pincus \& Klein}, 2000). Taking into account joint subgrid-scale distributions of just two moments for two species complicates the math considerably (\textit{Larson \& Griffin,} 2012).

An alternative to parameterizing convection within grid scales O(100 km) is the multiscale modeling framework (\textit{Randall et al}., 2003), in which two-dimensional cloud-resolving models are embedded within each GCM column. While some convection-related aspects of the global circulation are treated well by such an approach, like the more traditional approach to GCM cloud parameterization, the pervasive and climatologically important regime of shallow marine convection is poorly represented in such models (e.g., \textit{Marchand \& Ackerman}, 2010). (Explicit resolution of such clouds requires horizontal resolution O(100 m) and vertical resolution O(10 m).) Avoiding issues related to embedding 2-D slices within GCM columns (problems including how to orient the slices and shortcomings including the use of periodic lateral boundary conditions for each embedded 2-D sub-model), the most expensive, yet perhaps straightforward approach to climate modeling is the Earth Simulator, a global cloud-resolving model with simulations run on a 3.5-km horizontal grid (\textit{Satoh et al}., 2008). The computational demands of such an approach are vast, with order 10\textsuperscript{19} grid cells in such a model. Even such a brute-force approach falls far short of the grid resolution required to explicitly simulate shallow marine convection, however. It is safe to say that even on the most powerful computing platforms that global simulations of Earth will be saddled with cloud and convective parameterizations for the foreseeable future.

\subsection{Exoplanet Clouds}

In extreme contrast to the situation for Earth outlined above, there are currently no observational constraints for atmospheres of terrestrial exoplanets that would provide information about what kind of clouds may have to be considered for a particular exoplanet. Available observables are confined to basic planetary parameters, like radius and orbital inclination (if a transit event can be observed), planetary mass (from the radial velocity method), orbital eccentricity and distance, and additionally the type of the central star. Consequently, we are faced with the difficult problem of modeling clouds in an environment without actually knowing any further details.

Cloud formation can only be treated theoretically in compliance with the chemical composition of the atmosphere because it determines the condensing species forming a cloud. Considering how diverse atmospheres of terrestrial exoplanets can be expected to be, the self-consistent modelling of cloud formation in such atmospheres without observational constraints or theoretical predictions is somewhat ambiguous. The composition of a terrestrial planet cannot be easily deduced from simple theoretical arguments. In the case of a planet which has lost its primordial atmosphere the atmospheric composition is determined by the combination of the outgassed chemical species from the planet's interior and the volatiles delivered by impacts of asteroids and comets and will therefore depend on the planet's mantle composition, physical processes in the planetary interior, and the composition and sizes of impactors. Another factor with a huge impact on the atmospheric composition is also the possible existence of a biosphere that interacts chemically with the atmosphere.

The long-term evolution of the atmospheric composition, such as that resulting from the carbonate-silicate cycle as on Earth, depends also on the occurrence of plate tectonics. To date, the only known planet with plate tectonics is Earth. It is currently not quite well understood under which conditions a planetary crust will start plate tectonics and how this process is maintained over an extended period of time. This is especially true for more massive terrestrial planets like super-Earths, where there is much controversy regarding plate tectonics (see e.g. \textit{Valencia et al. }2007, \textit{O'Neill et al. }2007).

Other important processes determining the atmospheric composition are the escape mechanisms of atmospheric gas to space. While thermal escape is a function of the atmospheric temperature, planetary mass, and the molecular weight, the non-thermal escape processes (erosion of the atmosphere by a stellar wind, for example) are much more complicated (e.g. \textit{Lammer et al. }2008 or \textit{Tian}, this volume). They not only depend on the activity of the central star-{}-which itself is a function of the stellar type and its particular stellar evolution-{}-but also on the possible existence of a planetary magnetic field (linked to the rotation rate of the planet), which can protect the planetary atmosphere against loss processes.

In contrast to a low-mass planet like Earth, a more massive terrestrial super-Earth might retain a part of its primordial hydrogen dominated atmosphere, partly enriched by volcanic outgassing or additional external delivery. Such atmospheres may be vastly different from those known within our solar system. A discussion of possible atmospheric compositions can be found in \textit{Seager \& Deming }(2010).

The secondary atmosphere of an Earth-like planet (Earth-like with respect to the chemical composition of the planetary mantle) would most likely be rich in H\textsubscript{2}O and CO\textsubscript{2} (\textit{Schaefer et al. }2012). Therefore, clouds composed of these species are of prime interest for habitable Earth-like planets. Cooler atmospheres can also contain significant amounts of CH\textsubscript{4} and CH\textsubscript{3}, or SO\textsubscript{2} in case of high temperatures (\textit{Schaefer et al. }2012).

In this environment effective cloud models must match the problem at hand. For example highly sophisticated terrestrial cloud microphysics and dynamics models are not required in order to ascertain the range of plausible albedos for a hypothetical Earth-like terrestrial planet. However more sophisticated approaches than simple ad-hoc models may be needed to interpret the colors of a directly imaged planet.

\subsection{Cloud Models for Terrestrial Exoplanets}

In principle the basic mathematical description of cloud microphysics in atmospheres of terrestrial exoplanets do not deviate from their solar system counterparts. The temporal and spatial evolution of the cloud particle size distribution can be described by means of a master equation (``general dynamic equation'') incorporating all relevant gain and loss processes. This includes nucleation, evaporation, sedimentation, coagulation/coalescence, diffusion, or hydrodynamical transport (see \textit{Pruppacher \& Klett}, 1996 or \textit{Lamb \& Verlinde} 2011). While the numerical solution of the master equation is still quite challenging it can be efficiently performed by the methods summarised in \textit{Williams \& Loyalka }(1991) or by applying more advanced methods (e.g. continuous and discontinuous Galerkin methods by \textit{Sandu \& Borden}, 2003).

Self-consistent modeling of the formation and temporal and spatial evolution of clouds is an unsolved problem. An ideal treatment would require a thorough knowledge of the state of the atmosphere, including its composition, the spatial distribution of chemical species, atmospheric temperature and dynamics, and the size distribution and density of potential cloud condensation nuclei and heterogeneous freezing nuclei. However, even in the terrestrial atmosphere the formation of ice at temperatures warmer than the homogeneous freezing temperature for water (about 233 K for typical drop sizes) is not well understood (e.g., \textit{Fridlind et al}., 2007, 2012); far less, if anything, is known about heterogeneous freezing and condensation nuclei in other atmospheres and would hardly be detectable by remote observations of terrestrial extrasolar planets. Lack of laboratory data to derive the necessary microphysical rates under atmospheric conditions more ``exotic'' than found in the solar system further complicates the problem of describing cloud formation in atmospheres of exoplanets.

Additionally, many atmospheric models for exoplanets are restricted to one spatial dimension and are often considered to be stationary, which makes a detailed description of cloud microphysics very difficult. While the climatic effects of clouds can be approximately treated in a one-dimensional model atmosphere, a consistent modelling of cloud formation would, in principle, require a three-dimensional dynamical atmospheric model as described in Section 4.2. In comparison to one-dimensional models, however, three-dimensional general circulation models contain many more free parameters and are very computationally intensive. Properties such as the surface orography, distribution of surface types (fractions of oceans and land mass), and the local distribution of chemical species (affected by volcanism for example) play a major role for the dynamics and chemistry of the atmosphere (\textit{Joshi}, 2003), and affect directly the formation and evolution of clouds. Since none of these detailed properties are known for a terrestrial exoplanet many additional assumptions about the planet and its atmosphere have to enter the calculations. On the other hand, three-dimensional models are the only opportunity to obtain information about the possible temporal variation and fractional coverage of clouds and their distribution throughout the atmosphere. Such results might be required to analyze transmission spectra of terrestrial exoplanetary atmospheres containing patchy clouds.

Given the aforementioned challenges due to the lack of observational constraints, clouds in atmospheres of terrestrial exoplanets are usually treated in a simplified way. The simplest method to account for the effects of clouds in an atmospheric model is a modified surface albedo. This approach has been widely used in the past (e.g., \textit{Kasting et al.} (1993), \textit{Segura et al.} (2003, 2005), or \textit{Grenfell et al.} (2007)). The surface albedo of these kind of models is modified to yield a specified surface temperature for a given reference scenario. For example, a common reference scenario is an Earth-like planet around the Sun at a known orbital distance of 1 AU. The surface albedo is then adjusted to obtain the mean surface temperature of Earth (288 K), thereby mimicking the climatic effects of clouds. This adjusted surface albedo value is then used in all subsequent model calculations, assuming that the net effect of the clouds is invariant from changes of the atmospheric conditions, or type of central star. This approach makes no assumptions about the physical nature of the clouds, their composition, size, or optical properties but instead estimates their effects based on the original tuning to surface temperature. While a modified surface albedo can crudely describe the climatic effect of clouds, it cannot simulate their impact on the planetary spectra.

A more detailed approach is to consider model scenarios where the properties of clouds can be assumed to be approximately known. For a completely Earth-like planet, one could expect that the properties of clouds in such an atmosphere would closely resemble those found on Earth. This approach was used by \textit{Kitzmann et al.} (2010) to study the impact of mean Earth water clouds in the atmospheres of Earth-like extrasolar planets. In contrast to a modified surface albedo, the wavelength-dependent optical properties of the cloud particles are explicitly taken into account to study their effects for different incident stellar spectra and other parameters. Thus, in addition to the influence on the atmospheric and surface temperatures, their impact on the planetary spectra can also be studied in detail using this modelling approach (e.g., \textit{Kaltenegger et al.,} 2007). If the properties of the clouds are not known (e.g., for a CO\textsubscript{2} cloud in a thick CO\textsubscript{2} dominated atmosphere of a super-Earth) parameter studies can be performed, varying the cloud properties over a wide range to estimate the possible effect of clouds. This approach has been used by \textit{Forget \& Pierrehumbert} (1997) for CO\textsubscript{2} ice clouds in an early Martian atmosphere, for example.

More detailed treatments of cloud formation in exoplanetary atmospheres include simplified air parcel and vertically resolved one-dimensional cloud models based on those originally developed for the Earth atmosphere. Such models can be used to determine mean cloud properties under different atmospheric conditions (see \textit{Neubauer et al.} (2011, 2012) for several cloud species (e.g. H\textsubscript{2}O and H\textsubscript{2}SO\textsubscript{4}) or \textit{Zsom et al.} (2012) for water (droplet and ice) clouds, for example). However, these more detailed descriptions of cloud formation already need to include many additional assumptions, such as the distribution of cloud condensation nuclei, which strongly influence the resulting cloud properties.

While cloud models for terrestrial exoplanets so far lack many of the sophisticated and detailed cloud microphysics needed to reproduce the complicated cloud structures known from Earth observations they nonetheless offer an important first-order estimate of cloud effects in exoplanetary atmospheres. One of the largest uncertainties for one-dimensional models is the treatment of fractional clouds. Unless the atmosphere is globally supersaturated, thus resulting in a completely cloud-covered planet, the fraction of the atmosphere where clouds are present has to be introduced as a free parameter. Thus, one-dimensional models are incapable of organically describing planets with fractional cloud-cover.

Although one-dimensional models are commonly employed for many terrestrial exoplanet applications, three-dimensional models have also been used. For the terrestrial super-Earth Gliese 581d, \textit{Wordsworth et al. }(2011) adapted a Mars global circulation model that included a simplified treatment of CO\textsubscript{2} ice cloud formation. This microphysical model assumes a certain size and number density of cloud condensation nuclei and equally distributes the condensable material among them, accounting also for the sedimentation and hydrodynamical transport of the cloud particles within the atmosphere. The corresponding formation of CO\textsubscript{2} clouds lead to an increase of the surface temperature of Gliese 581d in their model calculations. The same approach was also used by \textit{Wordsworth et al. }(2010) in a one-dimensional atmospheric model for the same planet.

\subsection{Giant Planet Cloud Models}

Unlike the case for Earthlike planets where water clouds are the greatest concern, a wide variety of species may condense in the hydrogen-helium dominated atmospheres of giant planets. \textit{S\'anchez-Lavega et al.} (2004) review the standard framework for cloud formation in giant planets. Homogeneous condensation occurs when the partial pressure of a species in the gas phase exceeds its saturation vapor pressure at a given temperature in the atmosphere. \textit{S\'anchez-Lavega et al.} tabulate the vapor pressures for many relevant species. Other expressions for additional species can be found in \textit{Ackerman \& Marley} (2001) and \textit{Morley et al.} (2012). Curves tracing the set of pressure and temperature conditions at which a given species condenses assuming equilibrium chemistry (``condensation equilibrium curves'') are shown for many species in Figure 3 and a schematic illustration of the resulting cloud decks is shown in Figure 4.

Once a cloud layer forms the condensate is removed from the overlying atmosphere and thus is no longer available to react at lower temperatures higher in the atmosphere. Thus the calculation of the chemical equilibrium state for the atmosphere must account for the presence of the cloud. Such a ``condensation chemistry'' is distinct from equilibrium chemistry calculations in which the condensate remains in communication with the gas phase and is available for reaction at lower temperatures. Condensation chemistry is discussed in detail by \textit{Fegley \& Lodders} (1996), \textit{Lodders \& Fegley} (2002), and \textit{Visscher et al.} (2006) as well as by \textit{Burrows \& Sharp} (1999) in the context of brown dwarf models and \textit{Sudarsky et al. }(2003) for extrasolar giant planets. The schematic Figure 4 accounts for condensation chemistry. If iron were not sequestered into a deep cloud layer in Jupiter's atmosphere the Fe would react with gaseous H\textsubscript{2}S to form FeS, thus removing H\textsubscript{2}S from the observable atmosphere, in contradiction to observations (\textit{Fegley \& Lodders,} 1996).

Once the cloud base pressure is found the challenge is to describe the variation in cloud particle sizes and number densities above this level. Early attempts to develop cloud models for use in giant solar system atmospheres included the work of \textit{Lewis} (1969), \textit{Rossow} (1978), and \textit{Carlson et al.} (1988). These and other early works are reviewed by \textit{Ackerman \& Marley }(2001). In this subsection we focus on more recent modeling approaches that are in use today, in particular the cloud models of \textit{Ackerman \& Marley} and of \textit{Helling} and collaborators.

\begin{figure}[h!]
  \centering
  \includegraphics[scale=1.0]{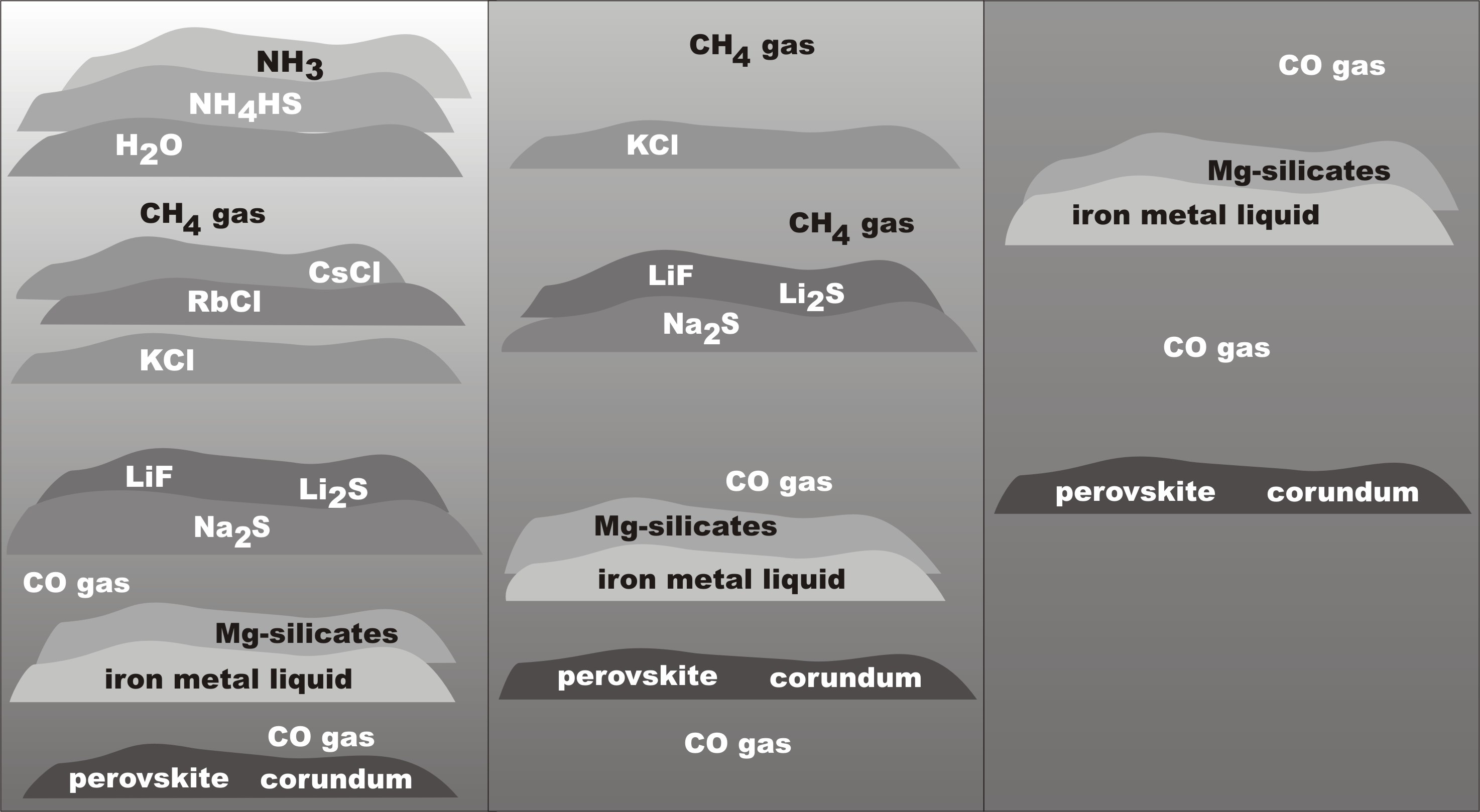}
  \caption{Schematic illustration (modified from \textit{Lodders} 2004) of cloud layers expected in extrasolar planet atmospheres based on consideration of equilibrium chemistry in the presence of precipitation. The three panels correspond roughly to effective temperatures \textit{T}\textsubscript{eff} of approximately 120 K (Jupiter-like, left), to 600 K (middle) to 1300 K (right). Note that with falling atmospheric temperature the more refractory clouds form at progressively greater depth in the atmosphere and new clouds composed of more volatile species form near the top of the atmosphere.}
\end{figure}

\subsubsection{Ackerman \& Marley}

Iron and silicates condense in the atmospheres of warm brown dwarfs (Figure 3) and these clouds must be accounted for in models of brown dwarf emergent spectra. Early modeling attempts for such clouds simply computed the mass of dust that would be found in chemical equilibrium for a given assumed initial abundance of gas (as if the gas were isolated at a given pressure and temperature from the rest of the atmosphere. The lower the atmospheric temperature the more dust that would be present. Atmospheric models following such a prescription (such as the ``DUSTY'' models of \textit{Allard et al.} 2001) adequately reproduced the near-infrared colors of the warmest brown dwarfs but predicted far too great of a dust load in cooler objects. Thus it was apparent that an accounting for sedimentation of grain particles was required. One approach used in the literature was to set a variable ``critical'' temperature for a given cloud such that cloud particles would only be found between cloud base and the specified temperature (\textit{Tsuji}, 2002). Another approach was to limit the cloud to be confined within an arbitrary distance, usually one scale-height, of cloud base. Both such approaches required the choice of an arbitrary particle size for the grains. The advantage of such approaches is that they are computationally very tractable for modeling and thus allow the exploration of a large parameter space. One disadvantage is that it is difficult to consider particle size effects and other complexities.

In order to allow for vertically-varying particle number densities and sizes a second approach was suggested by \textit{Ackerman \& Marley }(2001). In their formulation downward transport of particles by sedimentation is balanced by upwards mixing of vapor and condensate (either solid grains or liquid drops),

\begin{equation}
-K_{\mathrm{zz}}\frac{{\partial}q_t}{{\partial}z} - f_{\mathrm{sed}}w^* q_c = 0
\label{eq:cloud_formation_AM}
\end{equation}

where \textit{K}\textit{\textsubscript{zz}} is the vertical eddy diffusion coefficient, \textit{q}\textit{\textsubscript{t}} is the mixing ratio of condensate and vapor, \textit{q}\textit{\textsubscript{c}} is the mixing ratio of condensate, \textit{w}* is the convective velocity scale, and \textit{f}\textit{\textsubscript{sed}} is a dimensionless parameter that describes the efficiency of sedimentation. The solution of this equation allows computation of a self-consistent variation in condensate number density and particle size with altitude above an arbitrary cloud base.

In their model the cloud base is found by determining at which point in the atmosphere the local gas abundance exceeds the local condensate saturation vapor pressure  $P_{v,a}$ at which point the atmosphere becomes saturated. In cases where the formation of condensates does not proceed by homogeneous condensation an equivalent vapor pressure curve is computed, as described by \textit{Morley et al. }(2012).

The \textit{Ackerman \& Marley} cloud model has the advantage of not requiring knowledge of microphysical processes to compute particle sizes. For a given sedimentation efficiency clouds are simply assumed to have grown large enough to provide the required downward mass flux that balances Equation \eqref{eq:cloud_formation_AM}. Since the solution is numerically rapid a large number of models can be computed and compared with data within a tractable amount of time. Sample model temperature-pressure profiles along with equilibrium condensation curves are illustrated in Figure~3.

Considering the simplicity of this approach, the model has fared fairly well in comparisons with data. \textit{Stephens et al. }(2009) for example compared model spectra for L and T dwarfs to a large database of near- to mid-infrared spectra. They found that cloudy L dwarfs can generally be well fit by clouds computed with \textit{f}\textsubscript{sed} = 1 to 2 while early T dwarfs, which exhibit thinner clouds, are better fit with \textit{f}\textsubscript{sed} = 3 to 4. The model thus provides a framework for describing mean global clouds in a 1-dimensional sense, but the model lacks the ability to explain why the sedimentation efficiency might change at effective temperatures around 1200 K, where the near-infrared spectra of L dwarfs evolves over a small temperature range.

\textit{Marley et al}. (2010) considered the effect of partial cloudiness on L dwarf spectra computed with the \textit{Ackerman \& Marley }(2001) cloud model. Their method assumed that clear and cloudy columns of atmosphere had the same temperature profile and together emitted the flux corresponding to a specified effective temperature. They found that partially cloudy L dwarfs would have emergent spectra comparable to standard models with homogeneous cloud cover but with larger values of \textit{f}\textsubscript{sed}. Thus a dwarf with 50\% clear skies and 50\% cloudy skies with \textit{f}\textsubscript{sed} = 2 ends up with a model spectrum comparable to that of a homogeneous cloud cover with \textit{f}\textsubscript{sed~}= 4.

\subsubsection{Helling and Collaborators}

The most extensive body of work on cloud formation in giant exoplanet and brown dwarf atmospheres has been undertaken by Helling and her collaborators (\textit{Helling \& Woitke,} 2006; \textit{Helling et al., }2008; \textit{Witte et al.,} 2009, 2011; \textit{de Kok et al.,} 2011) who follow the trajectory of seed particles from the top of their model atmospheres as they sink downwards. The seeds grow and accrete condensate material as they fall, resulting in ``dirty'' or compositionally layered grains. This work extends the dust moment method from \textit{Gail et al.} (1984) and \textit{Gail \& Sedlmayr} (1988). It accounts for the microphysics of grain growth given these conditions and available relevant laboratory data. Particle nucleation is explicitly computed, taking into account barriers to grain formation.

Because condensation is envisioned in this framework to proceed downwards from the top of the atmosphere rather than upwards from the deep atmosphere and because grains are allowed to interact with the surrounding gas, the cloud composition predicted by the Helling approach differs substantially from that employed in the \textit{Ackerman \& Marley} framework. An example is shown in Figure 5 for a model brown dwarf (log \textit{g} = 5 in CGS units, \textit{T}\textsubscript{eff} = 1600 K). Here \textit{Helling et al.} (2008) predict that in addition to the usual Fe, MgSiO\textsubscript{3} and Mg\textsubscript{2}SiO\textsubscript{4} cloud layers, additional condensates including SiO\textsubscript{2} and MgO will form. These latter species are not predicted by equilibrium condensation for a cooling gas mixed upwards from higher temperature and pressure conditions The TiO\textsubscript{2} cloud seeds arising at the model at the top of the atmosphere in this approach is also evident.

In fact the presence of these initial TiO\textsubscript{2 }seed particles at the top of the model atmosphere in the \textit{Helling} framework deserves some discussion. In the equilibrium chemistry condensation framework, Ti-bearing condensates (e.g., CaTiO\textsubscript{3}) would form much deeper in the atmosphere as a gas parcel rises vertically and cools. Precipitation of such particles would remove the condensate from the gas phase and notable amounts of refractory TiO\textsubscript{2} seed particles would not be expected to arrive at the top of the atmosphere. Conversely in the \textit{Helling} conceptual framework the formation of CaTiO\textsubscript{3} at the expected equilibrium cloud base is thought to be kinetically inhibited (since multiple collisions of molecules would be required) resulting in the refractory TiO\textsubscript{2 }clusters being mixed further upwards to ultimately seed condensation in the cooler upper reaches of the atmosphere. Thus the \textit{Helling} approach fundamentally assumes that vertical mixing timescales, at least in localized columns, are faster than condensation timescales. As the seeds eventually fall from the top of the atmosphere they then encounter other condensable molecules which are likewise assumed not to have not been cold trapped below and the seeds then accrete these species and grow. The model iterates to find a self-consistent solution. This top-down approach thus differs from most of the other cloud modeling approaches discussed in the literature which generally conceive of a condensation sequence operating from the depths of the atmosphere upwards with species sequentially condensing, as conceptually shown in Figure 4. Atmospheric mixing by breaking gravity waves (Freytag et al. 2010) might provide a mechanism to stir the atmosphere sufficiently to deliver the seed particles.

Because of the complexity of the computational approach required to compute cloud properties in this framework there have been fewer comparisons between model spectra computed with the \textit{Helling} clouds and data than has been the case with the \textit{Ackerman \& Marley} cloud model. Some direct comparisons between different cloud models are shown by \textit{Helling et al. }(2008). Ultimately only a thorough comparison of the predictions of all cloud modeling schools and data will be required to establish which conceptual framework is a better approximation over which ranges of conditions. An application of the Helling framework to the clouds of Jupiter would be enlightening.

\begin{figure}[h!]
  \centering
  \includegraphics[scale=0.5]{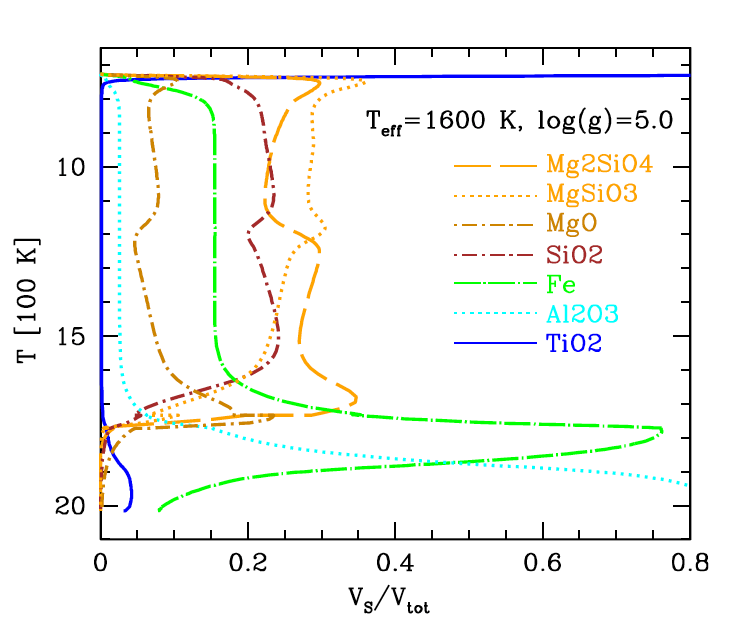} 
  \caption{Composition of atmospheric cloud layers for a T\textsubscript{eff }= 1600 K, log(g)=5 brown dwarf as computed by the dust model of Helling and collaborators (\textit{Helling et al.} 2008). The vertical axis is atmospheric temperature with the top of the atmosphere to the top of the figure. The horizontal axis gives the relative volumes \textit{V}\textsubscript{s} of each dust species indicated by line labels as a ratio to the total dust volume \textit{V}\textsubscript{tot}. Unlike the \textit{Ackerman \& Marley} condensation equilibrium approach, this model predicts that MgO and SiO\textsubscript{2} are important condensates along with the TiO\textsubscript{2} seed particles that are formed at the top of the model.}
\end{figure}

\subsubsection{Other Approaches}

More simplified approaches have also been taken for cloud models, such as specifying particle sizes and cloud heights. \textit{Sudarsky et al.} (2000; 2003) computed model exoplanet albedo spectra given various particle size and cloud height assumptions. In particular they utilized the \textit{Deirmendjian} (1964) size distribution and explored the effects of changing mean cloud particle sizes and widths of the size distribution. \textit{Sudarsky et al.} (2000) also considered the effects of various plausible photochemical hazes on giant planet albedo spectra.

\textit{Cooper et al. }(2003) employed the timescale comparison framework pioneered by \textit{Rossow} (1978) to compute cloud models for brown dwarfs and extrasolar giant planets. In this approach various timescales for particle nucleation, growth, and sedimentation are compared to derive expected condensate particle sizes. As discussed by \textit{Ackerman \& Marley }(2001), a difficulty with the Rossow method is that some of the critical timescales depend upon unknown factors, particularly the assumed supersaturation. Nevertheless the \textit{Cooper et al.} model provides a useful survey of likely particle sizes for species of interest expected under various combinations of gravity and effective temperature. For example, in agreement with most of the other cloud models \textit{Cooper et al. }predict typical silicate grain sizes in the range of 10 to 200 \textgreek{m}m.

Tsuji and collaborators (\textit{Tsuji }2002, 2005; \textit{Tsuji et al.} 2004) have computed brown dwarf models by specifying cloud-top and cloud-base temperatures. For the directly imaged planets \textit{Currie et al. }(2011), \textit{Madhusudhan et al.} (2011), and \textit{Bowler et al.} (2010) employ a variety of approaches to specify cloud properties and explore parameter space. Approaches such as these offer the advantage of quickly exploring the phase space of possible models and establishing the effect of plausible cloud models on spectra. The lack of physical complexity in such models is offset by their useful ability to offer qualitative understanding of the effect of various condensate properties.

\subsection{Lessons Learned from Cloudy Brown Dwarfs}

Brown dwarfs-{}-hydrogen-helium rich objects with masses between about 12 and 80 times that of Jupiter (M\textsubscript{J})-{}-have been a proving ground for understanding the role of clouds in exotic atmospheres. This is because the class of brown dwarfs known as the L dwarfs have atmospheric temperatures in the regime in which iron and silicate grains condense from the gas phase. It is apparent from the available data that these refractory condensates do not form a pall of particles mixed through the entire atmosphere, but rather form discrete cloud layers. As such L dwarfs were the first objects outside the solar system for which a detailed description of clouds was required in order to interpret their emitted spectra.

There have been several comparisons of cloudy brown dwarf atmosphere models to observations. The most extensive to date are presented by \textit{Cushing et al. }(2008) and \textit{Stephens et al.} (2009). These authors compared atmosphere models of \textit{Saumon \& Marley }(2008) computed using the \textit{Ackerman \& Marley} (2001) cloud model to a variety of L- and T-type brown dwarf spectra from 0.8 to 15 \textgreek{m}m. For the L dwarfs and early T-dwarfs the cloudy models clearly did a better job reproducing the data than cloudless models. The tunable \textit{f}\textsubscript{sed }parameter, with typical values between 1 and 3, allowed sufficient dynamic range to generally reproduce most of the observed spectra. \textit{Witte et al. }(2011) meanwhile compare model spectra computed with the \textit{Helling et al. }(2008) cloud formulation to spectra of L-type brown dwarfs. Likewise their cloud model shows much better agreement with data than either cloudless or very cloudy models with no dust sedimentation. In all of these studies the matches between models and data are very good in many cases, but nevertheless there remain notable spectral mismatches and it is clear that a more sophisticated cloud model would be required to fit all objects.

The most important lesson learned from the campaign to model brown dwarfs may be that large grids of atmosphere models-{}-including a variety of cloud descriptions-{}-are required. Models should not be so complex that the creation of such grids are a challenge. In the next section we review the available exoplanet data relevant to clouds. As we will see the exoplanet data do not yet require large systematic model grids, but such models will undoubtedly be required as more data become available. Ideally more models will be available than has been the case for brown dwarfs, thereby permitting more systematic comparisons of various cloud modeling frameworks to large exoplanet datasets.

\section{Observations of exoplanet clouds}

\subsection{Transiting Planets}

\subsubsection{Transmission Spectra}

Perhaps the most convincing evidence of high altitude clouds or hazes on any extrasolar planet to date is found in the case of the transiting planet HD 189733b. This 1.1-Jupiter mass planet orbits a bright nearby K star and is thus an excellent target for detailed studies of its atmosphere. This planet is notable because its transit radius-{}-the apparent size of the planet as a function of wavelength-{}-follows a smooth power law as first measured by \textit{Pont et al.} (2008). Signatures of molecular or atomic absorption expected from a clear, solar composition atmosphere are almost absent, although Na is detected at high spectral resolution (\textit{Huitson et al}., 2012). Figure 6 illustrates the smooth variation in atmospheric transmission as measured from 0.3 through 1 \textgreek{m}m (\textit{Pont et al., }2008; \textit{Sing et al}., 2009; \textit{Gibson et al}., 2011). The red curve presents a pure Rayleigh scattering model while a gas opacity only model is also shown. With the possible exception of a spectral feature at 1.5 \textgreek{m}m the smooth variation of planet radius with wavelength extends to at least 2.5 \textgreek{m}m (\textit{Gibson et al., }2011).

\begin{figure}[h!]
  \centering
  \includegraphics[scale=0.3]{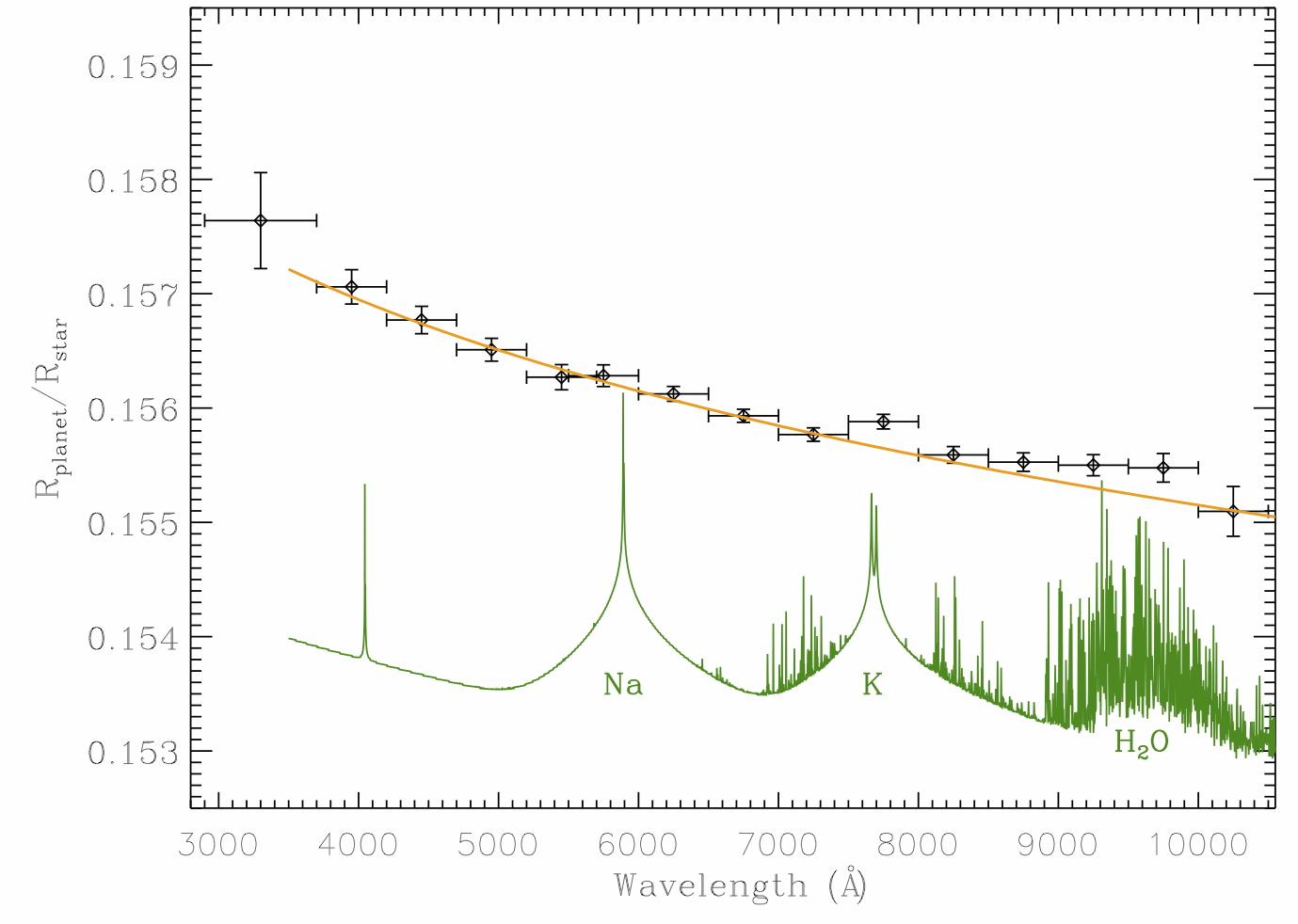}
	\caption{Observed transmission spectrum (points) for transiting hot Jupiter planet HD 189733b (\textit{Sing et al.} 2011). The width of the wavelength bin for each measurement is indicated by the x-axis error bars. The y-axis error bars denote the 1-\textgreek{s }error. The smooth curve denotes the prediction of a simple haze Rayleigh scattering-only model while the lower curve is a model for gaseous absorption only from \textit{Fortney et al. }(2010). Further details in \textit{Sing et al.} (2011). Figure courtesy J. Fortney.}
\end{figure}

The most natural explanation for the HD 189733b transmission spectrum is that a population of small, high albedo particles is present in the atmosphere at low pressures (\textit{Lecavelier des Etangs et al., }2008). In the terminology of Mie scattering this requires that the scattering efficiency is large compared to the absorption efficiency or $Q_{\mathit{abs}}{\ll}Q_{\mathit{scat}}$. \textit{Lecavelier des Etangs et al.} demonstrate that this in turn requires a material with an imaginary index of refraction that is small compared to the real index and suggest MgSiO\textsubscript{3} as a possible candidate. In the condensation chemistry framework (Section 3.4) however, silicates are expected to condense much deeper in the atmosphere and the mechanism by which small grains could be transported to the upper atmosphere has yet to be fully explored, although vigorous mixing is a likely explanation.

Another transiting planet for which particulate opacity may be important is GJ 1214b. This 6.5-Earth mass planet orbits an M star. Planetary structure models that fit the observed mass and radius of the planet can be found with either a thick hydrogen atmosphere comprising less than 3\% of the mass of the planet or a water rich planet surrounded by a steam atmosphere. Transmission spectra of a large scale height, hydrogen-dominated atmosphere (Figure 7) are predicted to exhibit well defined absorption bands while a water dominated atmosphere would have a much smaller scale height and consequently a very smooth transmission spectrum (\textit{Miller-Ricci \& Fortney,} 2010).

The observed flat transmission spectrum from the optical to perhaps 5 \textgreek{m}m (\textit{Bean et al}., 2010, 2011; \textit{Berta et al}., 2012; \textit{Desert et al., }2011) is consistent with a small scale-height atmosphere, thus apparently favoring the water-rich alternative. However high altitude clouds or hazes, as with HD 189733 b, could also be concealing atmospheric absorption bands if they lie at altitudes above 200 mbar (\textit{Bean et al}., 2010 and Figure 7). Since the nominal model pressure-temperature profile does not cross the condensation curve of any major species at solar abundance (although ZnS and KCl do condense around 1 bar), \textit{Bean et al. }proposed photochemical products as the likely source of the haze rather than clouds. However as shown in Figure 7, \textit{Morley et al. }(2013) find that very extended (\textit{f}\textsubscript{sed}=0.1) equilibrium sulphide clouds, including Na\textsubscript{2}S, in an atmosphere enriched in heavy elements can also attenuate the flux sufficiently to reproduce the data. Such small values of \textit{f}\textsubscript{sed} are seen in some terrestrial regimes (\textit{Ackerman \& Marley} 2001) but whether or not this would be plausible in the atmosphere of GJ 1214b remains to be investigated.

\textit{Miller-Ricci Kempton et al}. (2012) explored possible photochemical pathways in the atmosphere of this planet to explore possible mechanisms for haze production and found that second order hydrocarbons, including C\textsubscript{2}H\textsubscript{2}, C\textsubscript{2}H\textsubscript{4}, and C\textsubscript{2}H\textsubscript{6} are efficiently produced by UV photolysis of methane in the atmosphere. While their model did not explore the chemistry to hydrocarbons of order higher than C\textsubscript{2}H\textsubscript{6}, they argue that polymerization of the initial photochemical products are likely and that complex hydrocarbons including tholins and soots are likely to form. \textit{Morley et al. }(2013) demonstrate that such a hydrocarbon haze could well explain the flat transmission spectrum observed for this planet if the atmosphere is indeed hydrogen rich with efficient methane photolysis (with results very similar to the cloudy case shown in Figure 7). In an unpublished manuscript \textit{Zahnle et al}. (2009) argue that for solar composition atmospheres in general there is a range of atmospheric temperatures near 1000 K that would be expected to favor methane photolysis and the formation of higher order hydrocarbon soots or hazes. A definitive exploration of this point would require new generations of computer codes that can follow the fate of hydrocarbon species produced by photochemistry.

\begin{figure}[h!]
  \centering
  \includegraphics[scale=0.18]{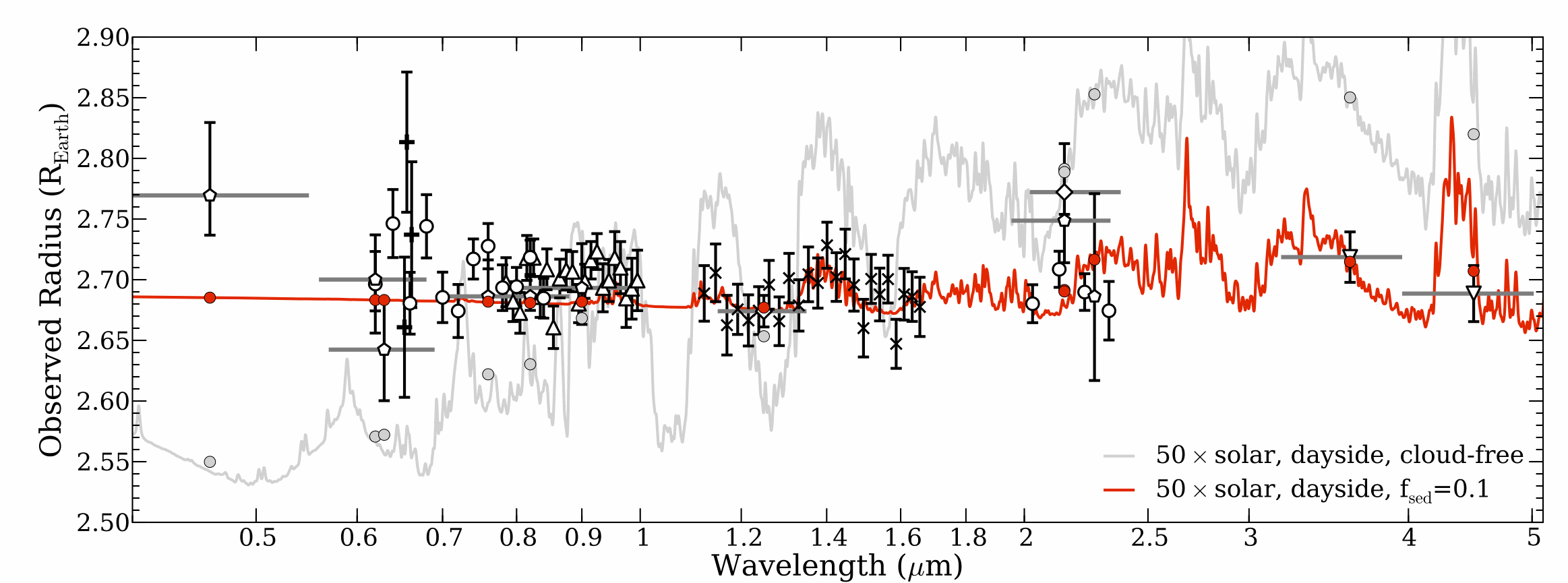}
	\caption{Observed (points) and model (lines) transit radius of GJ 1214b. Datapoints are from multiple observations as described in \textit{Morley et al.} (2013). Models are for a cloudless H\textsubscript{2}{}-He rich atmosphere with 50 times the solar abundance of heavy elements (grey) and the same atmosphere with the equilibrium abundance of clouds computed with \textit{f}\textsubscript{sed} = 0.1 (red). Figure modified from \textit{Morley et al.} (2013).}
\end{figure}

While it is still too soon to definitely characterize the atmospheres of either HD 189733b or GJ 1214b, it is apparent that cloud or haze opacity is likely important in at least some transiting planet atmospheres. The James Webb Space Telescope will obtain transit spectra of dozens of planets and characterize atmospheric absorbers as a function of planet mass and composition and the degree of stellar insolation. Such comprehensive studies will help to map out the conditions under which clouds and hazes are found.

\subsubsection{Transiting Planets in Reflected Light}

In addition to the light transmitted through the atmosphere of transiting planets, measurements have also been made of the the apparent brightness of the day side of transiting planets. By comparing the brightness of a transiting system immediately before and after a planet is occulted by its primary star the combination of light reflected and emitted by the planet can be measured. In the limit in which the planet does not emit but rather shines only by reflected light within the passband such a measurement provides the mean geometric albedo of the planet. However since such a measurement is most tractable for large planets orbiting close to their stars, in other words the hot Jupiter class of planets, thermal emission cannot generally be disregarded.

Upper limits on the geometric albedo have been placed on many planets (see the summary in \textit{Demory et al. }(2011)), generally finding the passband averaged geometric albedo to be less than about 30\% and in some cases much less (\textit{Kipping \& Spiegel,} 2011). Such a finding is not surprising for hot giant planet atmospheres (\textit{Marley et al., }1999; \textit{Sudarsky et al.,} 2000, 2003) since in the absence of a cloud layer most of the incident flux is absorbed rather than scattered. Despite these predictions as constraints became available from studies of transiting planets the preponderance of low albedos was often treated as surprising when compared to the bright albedo of Jupiter.

Persuasive evidence has been found, however, indicating that at least one hot Jupiter has a large geometric albedo that is likely attributable to a haze or cloud layer. Kepler-7b (\textit{Latham et al.}, 2010) has a geometric albedo averaged over the Kepler bandpass (423 to 897 nm, \textit{Koch et al.,} 2010) of  $0.32\pm 0.03$ (\textit{Demory et al.,} 2011). Such a high albedo is more typical of the cloudy solar system giants (e.g., \textit{Karkoschka, }1994) than a deep scattering atmosphere. \textit{Demory et al.} thus argue that the most likely explanation is that this particular planet has a bright photochemical haze or cloud layer, perhaps similar to that seen in the transit spectra of HD 189733b. More detailed modeling to constrain the properties of the scattering layer has not yet been done and to date transit spectra have not been obtained for Kepler 7b.

As albedos are measured for more planets with appropriate corrections for thermal emission, it may become apparent which particular combinations of planet mass, composition, and stellar incident flux (particularly including UV flux) conspire to produce high albedo planets.

\subsection{Directly Imaged Planets}

\subsubsection{Young Jupiters}

Giant planets start their lives in a warm, extended state with luminosities much greater than at later times after they have cooled. For this reason giant planets are easier to detect at young ages of a few hundred million years or less and several have already been detected, including 2M 1207b and the planets orbiting the A star HR 8799 (\textit{Marois et al.}, 2008; 2010). The near-infrared thermal emission of all of these objects (which have effective temperatures near 1000 K) points to the presence of substantial refractory cloud decks, most likely comprised of iron and silicate grains (e.g., \textit{Marois et al.}, 2008; \textit{Barman et al.}, 2011). Much of the literature on these objects has focused on constraining the properties of these clouds because in field brown dwarfs clouds have largely dissipated by 1000 K. \textit{Currie et al. }(2011), \textit{Madhusudhan et al.} (2011), and \textit{Bowler et al.} (2010) also all explored models for the directly imaged planets and agreed that clouds played a critical role in shaping their thermal emission at a lower effective temperature than is typical for more massive field brown dwarfs.

\textit{Marley et al.} (2012) also constructed model atmospheres for these planets and concurred that clouds are present in the atmospheres of HR 8799 b, c, and d. They applied the cloud model of \textit{Ackerman \& Marley} (2001) and found that they could reproduce much of the available data by using cloud parameters typically seen in warmer L dwarfs. They argue from mass balance considerations that mean cloud particle size likely varies inversely with  $\sqrt{g}$ where \textit{g} is the gravitational acceleration and that all else being equal the column optical depth of a cloud varies proportionately to  $\sqrt{g}$. The net result of these scaling relationships is that clouds in lower gravity objects tend to be similar to clouds found at warmer effective temperatures in higher gravity, more massive brown dwarfs. They find that as objects cool their spectra can be modeled by increasing the sedimentation efficiency, \textit{f}\textsubscript{sed} in Equation (1). The effective temperature at which \textit{f}\textsubscript{sed} begins to increase varies with gravity such that lower gravity objects begin to lose their refractory clouds at warmer effective temperatures than low mass objects.

At cooler effective temperatures near 600 K another set of condensates become important in field dwarfs. \textit{Morley et al.} (2012) have demonstrated that Na\textsubscript{2}S and other clouds (Figure 3) moderately redden the near-IR colors of late T type brown dwarfs. Once young giant planets in this effective temperature range are discovered it will be possible to test if the scaling relationships seen at higher effective temperature persist.

Ultimately understanding the nature of clouds in the warm, young giant planets hinges on understanding why the cloud clearing effective temperature varies as it does. In the next few years many more young giant planets are expected to be discovered by the upcoming GPI and SPHERE exoplanet surveys.

\subsubsection{Giants in Reflected Light}

No giant planet has yet been imaged in reflected light although such an observation is less technically difficult than imaging a terrestrial planet in reflected starlight and a number of space telescope missions have been proposed. As with terrestrial planets, clouds play a large role in affecting the reflection spectra of a giant planet (\textit{Marley et al., }1999; \textit{Sudarsksy et al.,} 2002; \textit{Cahoy et al., }2010).

One way of visualizing the effect of clouds in a giant planet atmosphere is to imagine a Jupiter twin lying at progressively closer distances to the sun (Figure 4). The optical geometric albedo spectrum of Jupiter generally consists of a bright continuum punctuated by methane absorption bands. The bright continuum is formed by scattering from gas and stratospheric hazes in the blue and the bright ammonia clouds at longer wavelengths. The methane bands are formed from the column of gas overlying the clouds. Below the ammonia cloud-tops lie cloud decks of NH\textsubscript{4}SH and H\textsubscript{2}O.

A Jupiter twin (with the same internal heat flux) lying closer to its star at 2AU would have a warmer atmosphere in which the ammonia and ammonium hydrosulphide clouds would not condense. Instead the planet would be covered by a bright global layer of water clouds which would give a very high albedo (Figure 1). At even closer distances to its star the atmosphere would be too warm for water clouds to form. Gas absorption thus overwhelms gaseous Rayleigh scattering and the planet becomes much darker (Figure 1) than the cloudless case.

\section{Additional Topics}

\subsection{Clouds of Low Porosity Aggregates}

Most of the discussion in this chapter has focused on cloud or haze particles as fully dense spheres. However fluffy or porous aggregate particles may very well form and such particles behave differently both as they interact with the atmosphere and with radiation. In this section we simplify the \textit{Ackerman \& Marley} (2001) approach to \textcolor[rgb]{0.4,0.4,0.4}{i}llustrate how low-porosity aggregates would behave. In this simplified version, we neglect the fraction of the condensible species present as vapor above the cloud base, so $q_t$=  $q_c$. We stress that this treatment is not a replacement for the complete model.

When the particles are too small to settle (see below) the condensate mixing ratio $q=q_c=\rho _c/\rho _g$ is constant and so has an infinite scale height $H_q$ (the condensate has the same scale height as the atmosphere). However, when the cloud is significantly settled such that the condensate mixing ratio scale height  $H_q{\ll}H,$ and if it is assumed that the vapor mixing ratio is negligible compared to the condensate ratio then Equation \eqref{eq:cloud_formation_AM} governing the vertical cloud distribution can be approximated as $K_{\mathit{zz}}\mathit{dq}/\mathit{dz}-v_fq=0$, where \textit{K}\textit{\textsubscript{zz}} = \textit{Lw*} is the vertical eddy diffusion coefficient, a property of macroscopic turbulence with typical lengthscale \textit{L} and large (energy-containing) eddy velocity \textit{w*; L} is usually taken to be the atmospheric scale height \textit{H,} although this might not be valid if the associated large eddy timescale \textit{L/w* }is much longer than the rotation period of the object (\textit{Schubert \& Zhang}, 2000) or the temperature profile is stable (\textit{Ackerman \& Marley} use an ad hoc stability correction). Then, for a settled cloud with uniform mass mixing ratio \textit{q} and effective thickness \textit{H}\textit{\textsubscript{q}}, the above equation is approximated by  $K_{\mathit{zz}}q/H_q-v_fq=0$, thus the cloud thickness is roughly  $H_q=Hw^* / v_f$. The scale height in the simplified exponential solution for condensate mass mixing ratio of \textit{Ackerman \& Marley }(2001) is essentially the same. That is, when the convective eddy velocity is much larger than the settling velocity, the cloud particles are not able to settle ( $f_{\mathit{sed}}=0$).

Particle settling velocities depend upon the dynamical regime. The settling velocity \textit{v}\textit{\textsubscript{f}} is the product of the local gravity\textit{ g} and the gas drag stopping time of the particle, \textit{t}\textit{\textsubscript{s}}\textit{.} \textit{Ackerman \& Marley }(2001), equation B2, contains a bridging expression \textit{\textgreek{b}} that covers both the so-called Stokes and Epstein drag regimes, in which \textit{r }is respectively greater than, or less than, the gas molecule mean free path  $l_m=m_{H_2}/\sigma _{H_2}\rho _g$, where  $m_{H_2}$ and  $\sigma _{H_2}$ are the mass and cross section of a hydrogen molecule and  $\rho _g$ is the gas density. In equations B1 and B3 of \textit{Ackerman \& Marley }(2001) the density contrast is essentially the particle density because\textit{ } $\rho {\gg}\rho _g$ for all applications of interest. So, \textit{v}\textit{\textsubscript{f}} is proportional to\textit{ \textgreek{r}}, and for particles with porosity  $\varphi $,  $\rho =\rho _s\left(1-\varphi \right)=\rho _sf$. In the Epstein regime (\textit{r/l}\textit{\textsubscript{m}}\textit{ {\textless} 1) }the expression for \textit{t}\textit{\textsubscript{s}} is very simple:  $t_s=r\rho /c\rho _g$ where \textit{c} is the sound speed. \textit{Cuzzi \& Weidenschilling }(2006) show how the stopping time in the Stokes regime (\textit{r/l}\textit{\textsubscript{m}}\textit{ {\textgreater} 1) }is essentially larger by a factor of \textit{r/l}\textit{\textsubscript{m}}.

We can thus show that for porous particles under local gravity \textit{g}, having the same mass as a particle with no porosity, $v_f=\mathit{gr}\rho /c\rho _g=gr_s\rho _s\left(1-\varphi \right)^{2/3}/c\rho _g$, and thus $H_q=Hw^*/v_f=H_{q\left(\mathit{solid}\right)}\left(1-\varphi \right)^{-2/3}$. Porous particles are lofted to greater heights than non-porous particles of the same mass, because of their slower settling velocity. In this sense they behave like smaller particles; however, their radiative behavior is not that of smaller particles, as noted above; porous particles in the regime $r/\lambda {\gg}1$ \textit{\ }indeed have large and wavelength-independent opacity.

\subsection{Polarization}

Polarization may provide an additional avenue for characterizing cloudy planets. Unlike the stellar radiation emitted by the central star, the light scattered by clouds will in general be polarized. Thus, investigating the polarized radiation scattered by clouds in contrast to the non-polarized stellar radiation may be an opportunity for characterizing cloudy exoplanetary atmospheres at short wavelengths in the future (see \textit{Stam} 2008).

Giant planets can also be polarized in the thermal infrared. In this case the required asymmetry in radiation emitted across the apparent planetary disk must be provided either by rotational flattening or irregularities in the global cloud cover. \textit{Marley \& Sengupta} (2011) investigated the former mechanism and \textit{de Kok et al.} (2011) the latter. Both sets of authors found that in the most favorable circumstances polarization fractions of a few percent were plausible and that in such cases polarization confirms the presence of a scattering condensate layer.

While polarization undoubtedly provides additional constraints on cloud particle sizes (e.g., \textit{Br\'eon \& Colzy}, 2000), condensate phase (e.g., \textit{van Diedenhoven et al}., 2012a), and even asymmetry parameter (\textit{van Diedenhoven et al., }2012b) and atmospheric structure, it may in practice be of limited value for studies of exoplanets. Within the solar system the most well known discovery attributable to a polarization measurement is the particle size of the clouds of Venus (\textit{Hansen \& Hovenier}, 1974). An imaging photopolarimeter carried on the Pioneer 10 and 11 spacecraft also constrained the vertical structure of Jupiter's clouds (\textit{Gehrels et al., }1974). However in general other techniques have proven more valuable. Especially given the low signal to noise and difficulty inherent in any measurement of an extrasolar planet, the value of further dividing the light into polarization channels must be weighed against other potential measurements (e.g., obtaining higher resolution spectroscopy).

\section{Conclusions}

As with the planetary atmospheres of solar system planets, clouds are expected to play major roles in the vertical structure, chemistry, and reflected and emitted spectra of extrasolar planets. That said, the most extensive experience to date with clouds outside of the solar system has been with the brown dwarfs. The presence of refractory clouds in L dwarfs and sulphide and salt clouds in late T dwarfs has been well established and a number of methods have been developed to model these clouds.

Compared to cloud modeling approaches within the solar system or particularly on Earth, exoplanet cloud models are still in their infancy. In many cases arbitrary clouds are employed that specify a range of plausible cloud properties that are usually sufficient to explore parameter space. More sophisticated efforts attempt to derive particle sizes, composition, and number density as a function of height through the atmosphere from a given set of assumptions will be needed once higher resolution, broad wavelength spectral data become available.

As of the time of this chapter's writing the best evidence for clouds in extrasolar planet atmospheres lies in the spectra of the planets orbiting the nearby A star HR 8799. Planets b, c, and d each have red near-infrared colors that are best explained by global refractory cloud decks that have persisted to lower effective temperatures than in higher mass field brown dwarfs. This persistence of clouds to lower effective temperatures for lower gravity objects continues a trend that has already been recognized among brown dwarfs.

Among the transiting planets there is convincing evidence for high altitude clouds-{}-or perhaps a photochemical haze-{}-in the atmosphere of HD 189733b. The transit spectra of this planet lacks the deep absorption bands expected for a clear atmosphere of absorbing gas but rather exhibits a smoothly varying absorption profile likely caused by small grain scattering. A second planet, GJ 1214b also has a transit spectrum lacking absorption features. In this case the spectrum may be attributable either to a small atmospheric scale height resulting from a high mean molecular weight composition or from clouds.

These early detections are like distant clouds seen at sunset presaging a coming storm. The GPI, SPHERE, and many other direct imaging planet searches are expected to discover dozens of young, self-luminous extrasolar giant planets over the coming decade (\textit{Oppenheimer \& Hinkley}, 2009; \textit{Traub \& Oppenheimer,} 2010). Many of these planets will exist in the effective temperature range in which clouds shape their emergent spectra. Meanwhile continued studies of transiting planets, particularly by the upcoming James Webb Space Telescope will probe the atmospheres of the transiting planets and test the hypothesis that hazes of photochemical origin are important in some regimes. Finally future space based coronagraphic telescopes may eventually image extrasolar giant and terrestrial planets in reflected light. All such efforts to discover and characterize extrasolar planets will hinge upon an understanding of the role clouds play in shaping the climate and atmospheres of planets.

\section*{References}

\bigskip

\noindent{{Abe, Y., Abe-Ouchi, A., Sleep, N. H., and Zahnle, K. J. (2011) Habitable Zone Limits for Dry Planets}\textit{{. Astrobiology}}{, }\textit{{11(5)}}{, 443-460.}}

\bigskip

\noindent{\color[rgb]{0.0,0.0,0.039215688}
\textcolor[rgb]{0.2,0.2,0.2}{Ackerman, A.S., \& M.S. Marley (2001) Precipitating Condensation Clouds in Substellar Atmospheres.}\textit{Astrophys. J.}\textcolor[rgb]{0.2,0.2,0.2}{, }\textit{\textcolor[rgb]{0.2,0.2,0.2}{556}}\textcolor[rgb]{0.2,0.2,0.2}{, 872-884. }}

\bigskip

\noindent{\color[rgb]{0.0,0.0,0.039215688}
Allard, F., P. H. Hauschildt, D. R. Alexander, A. Tamani, and A. Schweitzer (2001) The Limiting Effects of Dust in Brown Dwarf Model Atmospheres. \textit{Astrophys. J., 556}, 357-372}

\bigskip

\noindent{\color[rgb]{0.0,0.0,0.039215688}
Arnold, L., Gillet, S., Lardi\`ere, O., Riaud, P., and Schneider, J. (2002) A test for the search for life on extrasolar planets. Looking for the terrestrial vegetation signature in the Earthshine spectrum. \textit{Astron. \& Astrophys.}, \textit{392}, 231--237.}

\bigskip

\noindent{\color[rgb]{0.0,0.0,0.039215688}
Barker, H. W., Stephens, G. L., and Fu, Q. (1999) The sensitivity of domain-averaged solar fluxes to assumptions about cloud geometry. \textit{Q. J. Roy. Met. Soc., 125}, 2127-2152.}

\bigskip

\noindent{\color[rgb]{0.0,0.0,0.039215688}
Barman, T.S., Macintosh, B., Konopacky, Q.M., \& Marois, C. (2011) Clouds and Chemistry in the Atmosphere of Extrasolar Planet HR8799b. \textit{Astrophys. J., 733}, 65 }

\bigskip

\noindent{\color[rgb]{0.0,0.0,0.039215688}
Bean, J.L., Miller-Ricci Kempton, E., \& Homeier, D. (2010) A ground-based transmission spectrum of the super-Earth exoplanet GJ 1214b. \textit{Nature, 468}, 669-672. }

\bigskip

\noindent{\color[rgb]{0.0,0.0,0.039215688}
Bean, J.L., D\'esert, J.-M., Kabath, P., et al. (2011) The Optical and Near-infrared Transmission Spectrum of the Super-Earth GJ 1214b: Further Evidence for a Metal-rich Atmosphere. \textit{Astrophys. J., 743}, 92. }

\bigskip

\noindent\noindent{\color[rgb]{0.0,0.0,0.039215688}
Beckwith, S. V. W., Henning, T., Nakagawa, Y. (2000) Dust Properties and Assembly of Large Particles in Protoplanetary Disks. In \textit{Protostars and Planets IV} (V. Mannings, A.P. Boss, and S. S. Russell, eds.), p. 533. University of Arizona Press, Tucson.}

\bigskip

\noindent\noindent{\color[rgb]{0.0,0.0,0.039215688}
Berta, Z.K., Charbonneau, D., D\'esert, J.-M., et al. (2012) The Flat Transmission Spectrum of the Super-Earth GJ1214b from Wide Field Camera 3 on the Hubble Space Telescope. \textit{Astrophys. J.}, \textit{747}, 35 }

\bigskip

\noindent\noindent{\color[rgb]{0.0,0.0,0.039215688}
Blum, J. (2010) Dust growth in protoplanetary disks - a comprehensive experimental/theoretical approach. \textit{Research in Astronomy and Astrophysics}, \textit{10}, 1199-1214.}

\bigskip

\noindent{\color[rgb]{0.0,0.0,0.039215688}
Bohren, C. and D. R. Huffman (1983) Absorption and scattering of light by small particles, New York: Wiley}

\bigskip

\noindent{\color[rgb]{0.0,0.0,0.039215688}
Bony, S., \& Dufresne, J.-L. (2005) Marine boundary layer clouds at the heart of tropical cloud feedback uncertainties in climate models. \textit{Geophys. Res. Let.}, \textit{32}, 20806 }

\bigskip

\noindent{\color[rgb]{0.0,0.0,0.039215688}
Bowler, B.P., Liu, M.C., Dupuy, T.J., \& Cushing, M.C. (2010) Near-infrared Spectroscopy of the Extrasolar Planet HR 8799 b. \textit{Astrophys. J.,} \textit{723}, 850 }

\bigskip

\noindent{\color[rgb]{0.0,0.0,0.039215688}
Br\'eon, F., and S. Colzy (2000) Global distribution of cloud droplet effective radius from POLDER polarization measurements. \textit{Geophys. Res. Lett}.,\textit{ 27(24),} 4065--4068.}

\bigskip

\noindent{\color[rgb]{0.0,0.0,0.039215688}
Burrows, A., \& G. Orton (2010). Giant Planet Atmospheres and Spectra. In \textit{Exoplanets }(S. Seager, ed.), p. 419. Univ. Arizona Press, Tucson.}

\bigskip

\noindent{\color[rgb]{0.0,0.0,0.039215688}
Burrows, A., \& C.M. Sharp (1999) Chemical Equilibrium Abundances in Brown Dwarf and Extrasolar Giant Planet Atmospheres. \textit{Astrophys. J.,} \textit{512}, 843 }

\bigskip

\noindent{\color[rgb]{0.0,0.0,0.039215688}
Cahoy, K., Marley, M., Fortney, J. (2010) Exoplanet albedo spectra and colors as a function of planet phase, separation, and metallicity. \textit{Astrophys. J.,} \textit{724}, 189-214.}

\bigskip

\noindent{\color[rgb]{0.0,0.0,0.039215688}
Carlson, B.E., W.B. Rossow, \& G.S. Orton (1988) Cloud microphysics of the giant planets.}
{\color[rgb]{0.0,0.0,0.039215688}
\ \textit{J. of Atmospheric Sciences, 45}, 2066 }

\bigskip

\noindent{\color[rgb]{0.0,0.0,0.039215688}
Colaprete, A. and Toon, O. B. (2003) Carbon dioxide clouds in an early dense Martian atmosphere. \textit{J. Geophys. Res. (Planets)}, \textit{108}, 5025.}

\bigskip

\noindent{\color[rgb]{0.0,0.0,0.039215688}
Cooper, C., Sudarsky, D., Milsom, J., Lunine, J., Burrows, A. (2003) Modeling the formation of clouds in brown dwarf atmospheres. \textit{Astrophys. J.,} \textit{586}, 1320.}

\bigskip

\noindent{\color[rgb]{0.0,0.0,0.039215688}
Currie, T., A. Burrows, Y. Itoh, et al. (2011) A Combined Subaru/VLT/MMT 1-5 \textgreek{m}m Study of Planets Orbiting HR 8799: Implications for Atmospheric Properties, Masses, and Formation.}
{\color[rgb]{0.0,0.0,0.039215688}
\textit{Astrophys. J.,} \textit{729}, 128 }

\bigskip

\noindent{\color[rgb]{0.0,0.0,0.039215688}
Cushing, M.C., Marley, M.S., Saumon, D., Kelly, B., Vacca, W., Rayner, J., Freedman, R., Lodders, K., Roellig, T. (2008) Atmospheric Parameters of Field L and T Dwarfs. \textit{Astrophys. J.}, \textit{678}, 1372.}

\bigskip

\noindent{\color[rgb]{0.0,0.0,0.039215688}
Cuzzi, J. N. and R. C. Hogan (2003) Blowing in the wind: I. Velocities of Chondrule-sized Particles in a Turbulent Protoplanetary Nebula. \textit{Icarus}, \textit{164}, 127-138.}

\bigskip

\noindent{\color[rgb]{0.0,0.0,0.039215688}
Cuzzi, J. N. and Weidenschilling, S. (2006) Particle-Gas Dynamics and Primary Accretion. In \textit{Meteorites and the Early Solar System II} (D. S. Lauretta and H. Y. McSween Jr., eds.), pp. 353-381. University of Arizona Press, Tucson.}

\bigskip

\noindent{\color[rgb]{0.0,0.0,0.039215688}
de Kok, R.J., C. Helling, D.M. Stam, P. Woitke, \& S. Witte (2011) The influence of non-isotropic scattering of thermal radiation on spectra of brown dwarfs and hot exoplanets. \textit{Astron. \& Astrophys.}, \textit{531}, A67 }

\bigskip

\noindent{\color[rgb]{0.0,0.0,0.039215688}
Deirmendjian, D. (1964) Scattering and polarization of water clouds and hazes in the visible and infrared. \textit{Appl Optics IP}, \textit{3}, 187.}

\bigskip

\noindent{\color[rgb]{0.0,0.0,0.039215688}
Demory, B.-O., Seager, S., Madhusudhan, N., et al. (2011) The High Albedo of the Hot Jupiter Kepler-7 b. \textit{Astrophys. J. Let.}, \textit{735}, L12 }

\bigskip

\noindent{\color[rgb]{0.0,0.0,0.039215688}
D\'esert, J.-M., Bean, J., Miller-Ricci Kempton, E., et al. (2011) Observational Evidence for a Metal-rich Atmosphere on the Super-Earth GJ1214b. \textit{Astrophys. J. Let., 731}, L40 }

\bigskip

\noindent{\color[rgb]{0.0,0.0,0.039215688}
Dominik, C., Blum, J., Cuzzi, J. N., and Wurm, G. (2007) Growth of Dust as the Initial Step Toward Planet Formation. In \textit{Protostars and Planets V} (B. Reipurth, D. Jewitt, and K. Keil, eds.), pp.783-800. University of Arizona Press, Tucson.}

\bigskip

\noindent{\color[rgb]{0.0,0.0,0.039215688}
Dominik, C. and Tielens, A. G. G. M. (1997) The Physics of Dust Coagulation and the Structure of Dust Aggregates in Space. \textit{Astrophys. J.}, \textit{480}, 647.}

\bigskip

\noindent{\color[rgb]{0.0,0.0,0.039215688}
Draine, B. T. and Lee, H. M. (1984) Optical properties of interstellar graphite and silicate grains. \textit{Astrophys. J.}, \textit{285}, 89-108.}

\bigskip

\noindent{\color[rgb]{0.0,0.0,0.039215688}
Ehrenreich, D., Tinetti, G., Lecavelier Des Etangs, A., et al. (2006) The transmission spectrum of Earth-size transiting planets. \textit{Astron. \& Astrophys.}, \textit{448}, 379--393.}

\bigskip

\noindent{\color[rgb]{0.0,0.0,0.039215688}
Fegley, B. and Lodders, K. (1996) Atmospheric Chemistry of the Brown Dwarf Gliese 229B: Thermochemical Equilibrium Predictions. \textit{Astrophys. J.}, \textit{472}, L37.}

\bigskip

\noindent{\color[rgb]{0.0,0.0,0.039215688}
Ferguson, J. W., Heffner-Wong, A., Penley, J. L., Barman, T. S., and D. R. Alexander (2007) Grain Physics and Rosseland Mean Opacities. \textit{Astrophys. J.,} \textit{666}, 261-266.}

\bigskip

\noindent{\color[rgb]{0.0,0.0,0.039215688}
Freytag, B., F. Allard, H.-G. Ludwig, D. Homeier, \& M. Steffen (2010) The role of convection, overshoot, and gravity waves for the transport of dust in M dwarf and brown dwarf atmospheres}
{\color[rgb]{0.0,0.0,0.039215688}
\textit{Astron. \& Astrophys.}, \textit{513}, A19 }

\bigskip

\noindent{\color[rgb]{0.0,0.0,0.039215688}
Fu, Q., Yang, P., Sun, W. B. (1998) An Accurate Parameterization of the Infrared Radiative Properties of Cirrus Clouds for Climate Models. \textit{J. Climate},\textit{ 11}, 2223--2237.}

\bigskip

\noindent{\color[rgb]{0.0,0.0,0.039215688}
Fu, Q. (2007) A New Parameterization of an Asymmetry Factor of Cirrus Clouds for Climate Models. \textit{J. Atmos. Sci., 64}, 4140--4150.}

\bigskip

\noindent{\color[rgb]{0.0,0.0,0.039215688}
Forget, F. and Pierrehumbert, R. T. (1997) Warming Early Mars with Carbon Dioxide Clouds That Scatter Infrared Radiation. \textit{Science}, \textit{278}, 1273.}

\bigskip

\noindent{\color[rgb]{0.0,0.0,0.039215688}
Fortney, J., Shabram, M., Showman, A., Lian, Y., Freedman, R., Marley, M., Lewis, N. (2010) Transmission Spectra of Three-Dimensional Hot Jupiter Model Atmospheres. \textit{Astrophys. J., 709}, 1396-1406.}

\bigskip

\noindent{\color[rgb]{0.0,0.0,0.039215688}
Fridlind, A.M., A.S. Ackerman, G. McFarquhar, G. Zhang, M.R. Poellot, P.J. DeMott, A.J. Prenni, and A.J. Heymsfield (2007) Ice properties of single-layer stratocumulus during the Mixed-Phase Arctic Cloud Experiment (M-PACE): Part II, Model results, \textit{J. Geophys. Res., 112, }D24202.}

\bigskip

\noindent{\color[rgb]{0.0,0.0,0.039215688}
Fridlind, A.M., B. van Diedenhoven, A.S. Ackerman, A. Avramov, A. Mrowiec, H. Morrison, P. Zuidema, and M.D. Shupe (2012) A FIRE-ACE/SHEBA case study of mixed-phase Arctic boundary-layer clouds: Entrainment rate limitations on rapid primary ice nucleation processes, \textit{J. Atmos. Sci., 69,} 365-389.}

\bigskip

\noindent{\color[rgb]{0.0,0.0,0.039215688}
Gail, H.-P., Keller, R., and Sedlmayr, E. (1984) Dust formation in stellar winds. I - A rapid computational method and application to graphite condensation. \textit{Astron. \& Astrophys.}, \textit{133}, 320-332}

\bigskip

\noindent{\color[rgb]{0.0,0.0,0.039215688}
Gail, H.-P. and Sedlmayr, E. (1988) Dust formation in stellar winds. IV - Heteromolecular carbon grain formation and growth. \textit{Astron. \& Astrophys.}, \textit{206}, 153-168}

\bigskip

\noindent{\color[rgb]{0.0,0.0,0.039215688}
Gehrels, T., D. Coffeen, M. Tomasko, et al. (1974) The Imaging Photopolarimeter Experiment on Pioneer 10. \textit{Science}, \textit{183}, 318-320. }

\bigskip

\noindent{\color[rgb]{0.0,0.0,0.039215688}
Gibson, N.P., F. Pont, \& S. Aigrain (2011) A new look at NICMOS transmission spectroscopy of HD 189733, GJ-436 and XO-1: no conclusive evidence for molecular features. \textit{Mon. Not. R. Astron. Soc.}, \textit{411}, 2199 }

\bigskip

\noindent{\color[rgb]{0.0,0.0,0.039215688}
Glandorf, D. L., Colaprete, A., Tolbert, M. A., and Toon, O. B. (2002) CO2 Snow on Mars and Early Earth: Experimental Constraints. \textit{Icarus}, \textit{160}, 66--72.}

\bigskip

\noindent{\color[rgb]{0.0,0.0,0.039215688}
Goldblatt, C. and Zahnle, K. (2011) Clouds and the faint young sun paradox. \textit{Clim. Past}, \textit{7}, 203-220.}

\bigskip

\noindent{\color[rgb]{0.0,0.0,0.039215688}
Grenfell, J. L., Stracke, B., von Paris, P., Patzer, B., Titz, R., Segura, A., and Rauer, H. (2007). The response of atmospheric chemistry on earthlike planets around F, G and K Stars to small variations in orbital distance. \textit{Planet. Space Sci.}, \textit{55}, 661--671.}

\bigskip

\noindent{\color[rgb]{0.0,0.0,0.039215688}
Hamdani, S., Arnold, L., Foellmi, C., et al. (2006) Biomarkers in disk-averaged near-UV to near-IR Earth spectra using Earthshine observations. \textit{Astron. \& Astrophys.}, \textit{460}, 617--624.}

\bigskip

\noindent{\color[rgb]{0.0,0.0,0.039215688}
Hansen, G. B. (1997) The infrared absorption spectrum of carbon dioxide ice from 1.8 to 333 \textit{\textgreek{m}}m. \textit{J. Geophys. Res.}, \textit{102}, 21569--21588.}

\bigskip

\noindent{\color[rgb]{0.0,0.0,0.039215688}
Hansen, G. B. (2005) Ultraviolet to near-infrared absorption spectrum of carbon dioxide ice from 0.174 to 1.8 \textit{\textgreek{m}}m. \textit{J. Geophys. Res. (Planets)}, \textit{110}, E11003.}

\bigskip

\noindent{\color[rgb]{0.0,0.0,0.039215688}
Hansen, J. E. and Hovenier, J. W. (1974) Interpretation of the Polarization of Venus. \textit{J. Atmos. Sci.}, \textit{31}, 1137--1160.}

\bigskip

\noindent{\color[rgb]{0.0,0.0,0.039215688}
Hansen, J. E. and Travis, L. (1974) Light scattering in planetary atmospheres. \textit{Space Sci. Rev}, \textit{16}, 527-610.}

\bigskip

\noindent{\color[rgb]{0.0,0.0,0.039215688}
Hansen, J., Lacis, A., Rind, D., et al. (1984) Climate sensitivity: Analysis of feedback mechanisms. In \textit{Climate Processes and Climate Sensitivity} (J.E. Hansen and T. Takahashi, eds.), pp. 130-163. AGU Geophysical Monograph 29, Maurice Ewing Vol. 5. American Geophysical Union.}

\bigskip

\noindent{\color[rgb]{0.0,0.0,0.039215688}
Hart, M. H. (1979) Habitable Zones about Main Sequence Stars. \textit{Icarus}, \textit{37}, 351--357.}

\bigskip

\noindent{\color[rgb]{0.0,0.0,0.039215688}
Hearty, T., Song, I., Kim, S., and Tinetti, G. (2009) Mid-Infrared Properties of Disk Averaged Observations of Earth with AIRS. \textit{Astrophys. J.,} \textit{693}, 1763--1774.}

\bigskip

\noindent{\color[rgb]{0.0,0.0,0.039215688}
Helling, Ch. and Woitke, P. (2006) Dust in brown dwarfs. V. Growth and evaporation of dirty dust grains. \textit{Astron. \& Astrophys.}, \textit{455}, 325-338.}

\bigskip

\noindent{\color[rgb]{0.0,0.0,0.039215688}
Helling, Ch., et al. (2008) A comparison of chemistry and dust cloud formation in ultracool dwarf model atmospheres. \textit{Mon. Not. R. Astron. Soc.}, \textit{391}, 1854-1873.}

\bigskip

\noindent{\color[rgb]{0.0,0.0,0.039215688}
Hu, R., Cahoy, K., and Zuber, M. (2012) Mars atmospheric CO\textsubscript{2} condensation above the north and south poles as revealed by radio occultation, climate sounder, and laser ranging observations. \textit{J. Geophys. Res.} \textit{117}, 7002.}

\bigskip

\noindent{\color[rgb]{0.0,0.0,0.039215688}
Huitson, C.M., D.K. Sing, A. Vidal-Madjar, et al. (2012) Temperature-pressure profile of the hot Jupiter HD 189733b from HST sodium observations: detection of upper atmospheric heating. \textit{Mon. Not. R. Astron. Soc.}, \textit{422}, 2477}

\bigskip

\noindent{\color[rgb]{0.0,0.0,0.039215688}
Irvine, W. M. (1975) Multiple scattering in planetary atmospheres. \textit{Icarus, 25}, 175-204.}

\bigskip

\noindent{\color[rgb]{0.0,0.0,0.039215688}
Joshi, M. (2003) Climate Model Studies of Synchronously Rotating Planets. \textit{Astrobiology}, \textit{3}, 415--427.}

\bigskip

\noindent{\color[rgb]{0.0,0.0,0.039215688}
Kaltenegger, L. and Sasselov, D. (2011) Exploring the Habitable Zone for Kepler Planetary Candidates. \textit{Astrophys. J.}, \textit{736}, L25.}

\bigskip

\noindent{\color[rgb]{0.0,0.0,0.039215688}
Kaltenegger, L. and Traub, W. A. (2009) Transits of Earth-like Planets. \textit{Astrophys. J.}, \textit{698}, 519--527.}

\bigskip

\noindent{\color[rgb]{0.0,0.0,0.039215688}
Kaltenegger, L., Traub, W. A., and Jucks, K. W. (2007) Spectral Evolution of an Earth-like Planet. \textit{Astrophys. J.}, \textit{658}, 598--616.}

\bigskip

\noindent{\color[rgb]{0.0,0.0,0.039215688}
Karkoschka, E. (1994) Spectrophotometry of the jovian planets and Titan at 300- to 1000-nm wavelength: The methane spectrum. \textit{Icarus}, \textit{111}, 174-192.}

\bigskip

\noindent{\color[rgb]{0.0,0.0,0.039215688}
Kasting, J. F. (1988) Runaway and moist greenhouse atmospheres and the evolution of Earth and Venus. \textit{Icarus}, \textit{74}, 472--494.}

\bigskip

\noindent{\color[rgb]{0.0,0.0,0.039215688}
Kasting, J. F., Whitmire, D. P., and Reynolds, R. T. (1993) Habitable Zones around Main Sequence Stars. \textit{Icarus}, \textit{101}, 108--128.}

\bigskip

\noindent{\color[rgb]{0.0,0.0,0.039215688}
Kipping, D.M., \& D.S. Spiegel (2011) Detection of visible light from the darkest world. \textit{Mon. Not. R. Astron. Soc.}, \textit{417}, L88 }

\bigskip

\noindent{\color[rgb]{0.0,0.0,0.039215688}
Kitzmann, D., Patzer, A. B. C., von Paris, P., et al. (2010) Clouds in the atmospheres of extrasolar planets. I. Climatic effects of multi-layered clouds for Earth-like planets and implications for habitable zones. \textit{Astron. \& Astrophys.}, \textit{511}, A66.}

\bigskip

\noindent{\color[rgb]{0.0,0.0,0.039215688}
Kitzmann, D., Patzer, A. B. C., von Paris, P., Godolt, M., and Rauer, H. (2011a) Clouds in the atmospheres of extrasolar planets. II. Thermal emission spectra of Earth-like planets influenced by low and high-level clouds. \textit{Astron. \& Astrophys.}, \textit{531}, A62.}

\bigskip

\noindent{\color[rgb]{0.0,0.0,0.039215688}
Kitzmann, D., Patzer, A. B. C., von Paris, P., Godolt, M., and Rauer, H. (2011b) Clouds in the atmospheres of extrasolar planets. III. Impact of low and high-level clouds on the reflection spectra of Earth-like planets. \textit{Astron. \& Astrophys.}, \textit{534}, A63.}

\bigskip

\noindent{\color[rgb]{0.0,0.0,0.039215688}
Koch, D., et al. (2010) Kepler Mission Design, Realized Photometric Performance, and Early Science. \textit{Astrophys. J.}, \textit{713}, L79-L86.}

\bigskip

\noindent{\color[rgb]{0.0,0.0,0.039215688}
Lamb, D. and Verlinde, J. (2011) Physics and Chemistry of Clouds. Cambridge University Press}

\bigskip

\noindent{\color[rgb]{0.0,0.0,0.039215688}
Lammer, H., Kasting, J. F., Chassefi\`ere, E., Johnson, R. E., Kulikov, Y. N., Tian, F. (2008) Atmospheric Escape and Evolution of Terrestrial Planets and Satellites. \textit{Space Science Reviews}, \textit{139}, 399-436}

\bigskip

\noindent{\color[rgb]{0.0,0.0,0.039215688}
Lammer, H., et al. (2009) What makes a planet habitable? \textit{Astron. \& Astrophys. Rev.}, \textit{17}, 181--249.}

\bigskip

\noindent{\color[rgb]{0.0,0.0,0.039215688}
Larson, V. E., and Griffin, B. M. (2012) Analytic upscaling of a local microphysics scheme. Part I: Derivation. \textit{Q. J. Roy. Met. Soc.,} doi:10.1002/qj.1967.}

\bigskip

\noindent{\color[rgb]{0.0,0.0,0.039215688}
Latham, D., et al. (2010) Kepler-7b: A Transiting Planet with Unusually Low Density. \textit{Astrophys. J.,} \textit{713}, L140-L144.}

\bigskip

\noindent{\color[rgb]{0.0,0.0,0.039215688}
Lecavelier Des Etangs, A., F. Pont, A. Vidal-Madjar, D. Sing, D. (2008) Rayleigh scattering in the transit spectrum of HD 189733b. \textit{Astron. \& Astrophys.}, \textit{481}, L83-L86. }

\bigskip

\noindent{\color[rgb]{0.0,0.0,0.039215688}
Lesins, G., Chylek, P., and Lohmann, U. (2002) A study of internal and external mixing scenarios and its effect on aerosol optical properties and direct radiative forcing. \textit{J. Geophys. Res., 107(D10)}, 4094.}

\bigskip

\noindent{\color[rgb]{0.0,0.0,0.039215688}
Lewis, J.S. (1969) The clouds of Jupiter and the NH\textsubscript{3}{}-H\textsubscript{2}O and NH\textsubscript{3}{}-H\textsubscript{2}S systems. \textit{Icarus}, \textit{10}, 365-378. }

\bigskip

\noindent{\color[rgb]{0.0,0.0,0.039215688}
Lindzen, R. S., Chou, M.-D., and Hou, A. Y. (2001) Does the earth have an adaptive infrared iris? \textit{Bull. Amer. Meteor. Soc., 82}, 417-432.}

\bigskip

\noindent{\color[rgb]{0.0,0.0,0.039215688}
Liou , K.-N. (2002) An Introduction to Atmospheric Radiation. Elsevier}

\bigskip

\noindent{\color[rgb]{0.0,0.0,0.039215688}
Lodders, K. and Fegley, B. (2002) Atmospheric Chemistry in Giant Planets, Brown Dwarfs, and Low-Mass Dwarf Stars. I. Carbon, Nitrogen, and Oxygen, \textit{Icarus,} \textit{155}, 393.}

\bigskip

\noindent{\color[rgb]{0.0,0.0,0.039215688}
Manabe, S. and Wetherald, R. T. (1967) Thermal equilibrium of the atmosphere with a given distribution of relative humidity. \textit{J. Atmos. Sci., 24}, 241-259.}

\bigskip

\noindent{\color[rgb]{0.0,0.0,0.039215688}
M\"a\"att\"anen, A., Vehkam\"aki, H., Lauri, A., et al. (2005) Nucleation studies in the Martian atmosphere. \textit{J. Geophys. Res. (Planets)}, \textit{110}, 2002.}

\bigskip

\noindent{\color[rgb]{0.0,0.0,0.039215688}
Madhusudhan, N., A. Burrows, T. Currie (2011) Model Atmospheres for Massive Gas Giants with Thick Clouds: Application to the HR 8799 Planets and Predictions for Future Detections. \textit{Astrophys. J., 737}, 34 }

\bigskip

\noindent{\color[rgb]{0.0,0.0,0.039215688}
Marchand, R., and Ackerman, T. (2010) An analysis of cloud cover in multiscale modeling framework global climate model simulations using 4 and 1 km horizontal grids. \textit{J. Geophys. Res., 115,} D16207.}

\bigskip

\noindent{\color[rgb]{0.0,0.0,0.039215688}
Marley, M., Gelino, C., Stephens, D., Lunine, J., and Freedman, R. (1999) Reflected spectra and albedos of extrasolar giant planets. I. Clear and cloudy atmospheres. \textit{Astrophys. J.}, \textit{513}, 879-893.}

\bigskip

\noindent{\color[rgb]{0.0,0.0,0.039215688}
Marley, M., Saumon, D., and Goldblatt, C. (2010) A patchy cloud model for the L to T dwarf transition. \textit{Astrophys J.,} \textit{723}, L117-L121.}

\bigskip

\noindent{\color[rgb]{0.0,0.0,0.039215688}
Marley, M.S., \& S. Sengupta, (2011) Probing the physical properties of directly imaged gas giant exoplanets through polarization. \textit{Mon. Not. R. Astron. Soc.}, \textit{417}, 2874-2881. }

\bigskip

\noindent{\color[rgb]{0.0,0.0,0.039215688}
Marley, M.S., Saumon, D., Cushing, M., et al. (2012) Masses, Radii, and Cloud Properties of the HR 8799 Planets. \textit{Astrophys J., 754}, 135. }

\bigskip

\noindent{\color[rgb]{0.0,0.0,0.039215688}
Marois, C., B. Macintosh, T. Barman, et al. (2008) Direct Imaging of Multiple Planets Orbiting the Star HR 8799. \textit{Science,} \textit{322}, 1348 }

\bigskip

\noindent{\color[rgb]{0.0,0.0,0.039215688}
Marois, C., B. Zuckerman, Q.M. Konopacky, B. Macintosh, \& T. Barman (2010) Images of a fourth planet orbiting HR 8799. \textit{Nature}, \textit{468}, 1080. }

\bigskip

\noindent{\color[rgb]{0.0,0.0,0.039215688}
Miller-Ricci, E., \& J.J. Fortney (2010) The Nature of the Atmosphere of the Transiting Super-Earth GJ 1214b. \textit{Astrophys J. Let.,} \textit{716}, L74-L79.}

\bigskip

\noindent{\color[rgb]{0.0,0.0,0.039215688}
Miller-Ricci K., Zahnle K., and Fortney J.J. (2012) The Atmospheric Chemistry of GJ 1214b: Photochemistry and Clouds. \textit{Astrophys J., 745}, 3.}

\bigskip

\noindent{\color[rgb]{0.0,0.0,0.039215688}
Miyake, K., \& Y. Nakagawa (1993) Effects of particle size distribution on opacity curves of protoplanetary disks around T Tauri stars. \textit{Icarus}, \textit{106}, 20.}

\bigskip

\noindent{\color[rgb]{0.0,0.0,0.039215688}
Monta\~n\'es-Rodr\'iguez, P., Pall\'e, E., Goode, P. R., and Mart\'in-Torres, F. J. (2006) Vegetation Signature in the Observed Globally Integrated Spectrum of Earth Considering Simultaneous Cloud Data: Applications for Extrasolar Planets. \textit{Astrophys. J.}, \textit{651}, 544--552.}

\bigskip

\noindent{\color[rgb]{0.0,0.0,0.039215688}
Morley, C., J. Fortney, M. Marley, C. Visscher, D. Saumon, \& S. Leggett (2012) Neglected Clouds in T and Y Dwarf Atmospheres. \textit{Astrophys. J.,} \textit{756}, 172.}

\bigskip

\noindent{\color[rgb]{0.0,0.0,0.039215688}
Morley, C., J. Fortney, J., E. M.-R. Kempton, M. Marley, \& C. Visscher (2013) Quantitatively Assessing the Role of Clouds in the Transmission Spectrum of GJ 1214B. \textit{Astrophys. J.,} submitted.}

\bigskip

\noindent{\color[rgb]{0.0,0.0,0.039215688}
Morrison, H., and Gettelman, A. (2009) A new two-moment bulk stratiform cloud microphysics scheme in the Community Atmosphere Model, version 3 (CAM3). Part I: Description and numerical tests. \textit{J. Climate, 21,} 3642-3659.}

\bigskip

\noindent{\color[rgb]{0.0,0.0,0.039215688}
Neubauer, D., Vrtala, A., Leitner, J. J., Firneis, M. G., and Hitzenberger, R. (2011) Development of a Model to Compute the Extension of Life Supporting Zones for Earth-Like Exoplanets. \textit{Origins of Life and Evolution of the Biosphere}, \textit{41}, 545--552.}

\bigskip

\noindent{\color[rgb]{0.0,0.0,0.039215688}
Neubauer, D., Vrtala, A., Leitner, J. J., Firneis, M. G., and Hitzenberger, R. (2012) The life supporting zone of Kepler-22b and the Kepler planetary candidates: KOI268.01, KOI701.03, KOI854.01 and KOI1026.01. \textit{Planet. Space Sci.}, \textit{73}, 397-406}

\bigskip

\noindent{\color[rgb]{0.0,0.0,0.039215688}
O'Neill, C., Jellinek, A. M., Lenardic, A. (2007) Conditions for the onset of plate tectonics on terrestrial planets and moons. \textit{Earth and Planetary Science Letters}, \textit{261}, 20-32}

\bigskip

\noindent{\color[rgb]{0.0,0.0,0.039215688}
Oppenheimer, B. and Hinkley, S. (2009) High-Contrast Observations in Optical and Infrared Astronomy. \textit{Annu. Rev. Astron. Astrophys.}, \textit{47}, 253-289.}

\bigskip

\noindent{\color[rgb]{0.0,0.0,0.039215688}
Ossenkopf, V. (1991) Effective-medium theories for cosmic dust grains. \textit{Astron. \& Astrophys.}, \textit{251}, 210-219.}

\bigskip

\noindent{\color[rgb]{0.0,0.0,0.039215688}
Pall\'e, E., Zapatero Osorio, M. R., Barrena, R., et al. (2009) Earth's transmission spectrum from lunar eclipse observations. \textit{Nature}, \textit{459}, 814--816.}

\bigskip

\noindent{\color[rgb]{0.0,0.0,0.039215688}
Pierrehumbert, R. and Gaidos, E. (2011) Hydrogen Greenhouse Planets Beyond the Habitable Zone. \textit{Astrophys. J.}, \textit{734}, L13.}

\bigskip

\noindent{\color[rgb]{0.0,0.0,0.039215688}
Pincus, R. and Klein, S. A. (2000) Unresolved spatial variability and microphysical process rates in large-scale models. \textit{J. Geophys. Res., 105}, 27059-27065.}

\bigskip

\noindent{\color[rgb]{0.0,0.0,0.039215688}
Pollack, J. B. and Cuzzi, J. N. (1980) Scattering by nonspherical particles of size comparable to the wavelength - A new semi-empirical theory and its application to tropospheric aerosols. \textit{J. Atmos. Sci.}, \textit{37}, 868-881.}

\bigskip

\noindent{\color[rgb]{0.0,0.0,0.039215688}
Pollack, J.B., Hollenbach, D., Beckwith, S., et al. (1994) Composition and radiative properties of grains in molecular clouds and accretion disks. \textit{Astrophys. J.}, \textit{421}, 615-639.}

\bigskip

\noindent{\color[rgb]{0.0,0.0,0.039215688}
Pont, F., H. Knutson, R.L. Gilliland, C. Moutou, \& D. Charbonneau (2008) Detection of atmospheric haze on an extrasolar planet: the 0.55-1.05 \textgreek{m}m transmission spectrum of HD 189733b with the Hubble Space Telescope. \textit{Mon. Not. R. Astron. Soc.}, \textit{385}, 109 }

\bigskip

\noindent{\color[rgb]{0.0,0.0,0.039215688}
Pruppacher, H. and Klett, J. (1997) Microphysics of Clouds and Precipitation, 2nd ed. Kluwer Academic Publishers.}

\bigskip

\noindent{\color[rgb]{0.0,0.0,0.039215688}
Ramanathan, V. and Coakley, Jr., J. A. (1978) Climate modeling through radiative-convective models. \textit{Rev. Geophys. Space Phys., 14}, 465-489.}

\bigskip

\noindent{\color[rgb]{0.0,0.0,0.039215688}
Ramanathan, V. and Collins, W. (1991) Thermodynamic regulation of ocean warming by cirrus clouds deduced from observations of the 1987 El Nino. \textit{Nature, 351}, 27-32.}

\bigskip

\noindent{\color[rgb]{0.0,0.0,0.039215688}
Randall, D., Khairoutdinov, M., Arakawa, A., and Grabowski, W. (2003) Breaking the cloud parameterizations deadlock. \textit{Bull. Am. Meteorol. Soc., 84}, 1547--1564.}

\bigskip

\noindent{\color[rgb]{0.0,0.0,0.039215688}
Rossow, W.B. (1978) Cloud microphysics - Analysis of the clouds of Earth, Venus, Mars, and Jupiter. \textit{Icarus}, \textit{36}, 1-50. }

\bigskip

\noindent{\color[rgb]{0.0,0.0,0.039215688}
Satoh, M., Matsuno, T., Tomita, H., et al. (2008) Nonhydrostatic icosahedral atmospheric model (NICAM) for global cloud resolving simulations. \textit{J. Comp. Phys., 227}, 3486-3514.}

\bigskip

\noindent{\color[rgb]{0.0,0.0,0.039215688}
S\'anchez-Lavega, A., P\'erez-Hoyos, S., and Hueso, R. (2004) Clouds in planetary atmospheres: A useful application of the Clausius-Clapeyron equation. \textit{American J. of Physics}, \textit{72}, 767-774.}

\bigskip

\noindent{\color[rgb]{0.0,0.0,0.039215688}
Sandu, A. and Borden, C. (2003) A framework for the numerical treatment of aerosol dynamics. \textit{Appl. Numer. Math.}, \textit{45(4)}, 475--497.}

\bigskip

\noindent{\color[rgb]{0.0,0.0,0.039215688}
Saumon, D. and Marley, M. (2008) The evolution of L and T dwarfs in color-magnitude diagrams. \textit{Astrophys. J.}, \textit{689}, 1327-1344.}

\bigskip

\noindent{\color[rgb]{0.0,0.0,0.039215688}
Schaefer, L., Lodders, K., Fegley, B. (2012) Vaporization of the Earth: Application to Exoplanet Atmospheres. \textit{Astrophys. J.}, \textit{755}, 41}

\bigskip

\noindent{\color[rgb]{0.0,0.0,0.039215688}
Schmidt, G.A., et al. (2006) Present day atmospheric simulations using GISS ModelE: Comparison to in-situ, satellite and reanalysis data. \textit{J. Climate, 19,} 153-192.}

\bigskip

\noindent{\color[rgb]{0.0,0.0,0.039215688}
Schneider, S. H. (1972) Cloudiness as a global climatic feedback mechanism: The effects on the radiation balance and surface temperature of variations in cloudiness. \textit{J. Atmos. Sci., 29,} 1413-1422.}

\bigskip

\noindent{\color[rgb]{0.0,0.0,0.039215688}
Schneider, S. H. and Dickinson, R. E. (1974) Climate modeling. \textit{Rev. Geophys. Space Phys., 12}, 447-493.}

\bigskip

\noindent{\color[rgb]{0.0,0.0,0.039215688}
Schubert, G. and Zhang, K. (2000) Dynamics of Giant Planet Interiors. In \textit{From Giant Planets to Cool Stars}, ASP Conf. Series 212 (C. Griffith and M. Marley, eds.), 210-222. }

\bigskip

\noindent{\color[rgb]{0.0,0.0,0.039215688}
Seager, S. and Deming, D. (2010) Exoplanet Atmospheres, \textit{Annu. Rev. Astron. Astrophys.}, \textit{48}, 631.}

\bigskip

\noindent{\color[rgb]{0.0,0.0,0.039215688}
Segura, A., Kasting, J. F., Meadows, V., et al. (2005) Biosignatures from Earth-Like Planets Around M Dwarfs. \textit{Astrobiology}, \textit{5}, 706--725.}

\bigskip

\noindent{\color[rgb]{0.0,0.0,0.039215688}
Segura, A., Krelove, K., Kasting, J. F., et al. (2003) Ozone Concentrations and Ultraviolet Fluxes on Earth-Like Planets Around Other Stars. \textit{Astrobiology}, \textit{3}, 689--708.}

\bigskip

\noindent{\color[rgb]{0.0,0.0,0.039215688}
Selsis, F. (2004) The Atmosphere of Terrestrial Exoplanets: Detection and Characterization. In \textit{Extrasolar Planets: Today and Tomorrow} (J. Beaulieu, A. Lecavelier Des Etangs, and C. Terquem, eds.), p. 170. Astronomical Society of the Pacific Conference Series Vol. 321.}

\bigskip

\noindent{\color[rgb]{0.0,0.0,0.039215688}
Selsis, F., Kasting, J. F., Levrard, B., et al. (2007) Habitable planets around the star Gliese 581? \textit{Astron. \& Astrophys.}, \textit{476}, 1373--1387}

\bigskip

\noindent{\color[rgb]{0.0,0.0,0.039215688}
Sing, D.K., J.-M. D\'esert, A. Lecavelier Des Etangs, et al. (2009) Transit spectrophotometry of the exoplanet HD 189733b. I. Searching for water but finding haze with HST NICMOS. \textit{Astron. \& Astrophys.}, \textit{505}, 891-899. }

\bigskip

\noindent{\color[rgb]{0.0,0.0,0.039215688}
Sing, D.K., F. Pont, S. Aigrain, et al. (2011) Hubble Space Telescope transmission spectroscopy of the exoplanet HD 189733b: high-altitude atmospheric haze in the optical and near-ultraviolet with STIS. \textit{Mon. Not. R. Astron. Soc.}, \textit{416}, 1443 }

\bigskip

\noindent{\color[rgb]{0.0,0.0,0.039215688}
Stam, D. M. (2008) Spectropolarimetric signatures of Earth-like extrasolar planets. \textit{Astron. \& Astrophys.}, \textit{482}, 989--1007.}

\bigskip

\noindent{\color[rgb]{0.0,0.0,0.039215688}
Stephens, D., Leggett, S., Cushing, M., Marley, M., Saumon, D., Geballe, T., Golimowski, D., Fan, X., Noll, K. (2009) The 0.8-14.5 \textgreek{m}m Spectra of Mid-L to Mid-T Dwarfs: Diagnostics of Effective Temperature, Grain Sedimentation, Gas Transport, and Surface Gravity. \textit{Astrophys. J.,} \textit{702}, 154. }

\bigskip

\noindent{\color[rgb]{0.0,0.0,0.039215688}
Stognienko, R., Henning, Th., and Ossenkopf, V. (1995) Optical properties of coagulated particles. \textit{Astron. \& Astrophys.}, \textit{296}, 797-809.}

\bigskip

\noindent{\color[rgb]{0.0,0.0,0.039215688}
Sudarksy, D., Burrows, A., Pinto, P. (2000) Albedo and Reflection Spectra of Extrasolar Giant Planets. \textit{Astrophys. J.,} \textit{538}, 885-903.}

\bigskip

\noindent{\color[rgb]{0.0,0.0,0.039215688}
Sudarsky, D., Burrows, A., and Hubeny, I. (2003) Theoretical Spectra and Atmospheres of Extrasolar Giant Planets. \textit{Astrophys. J.,} \textit{588}, 1121-1148.}

\bigskip

\noindent{\color[rgb]{0.0,0.0,0.039215688}
Tinetti, G., J.P. Beaulieu, T. Henning, et al. (2012) EChO - Exoplanet Characterisation Observatory. \textit{Experimental Astronomy}, \textit{34}, 311-353.}

\bigskip

\noindent{\color[rgb]{0.0,0.0,0.039215688}
Tinetti, G., Meadows, V. S., Crisp, D., et al. (2006a) Detectability of Planetary Characteristics in Disk-Averaged Spectra. I: The Earth Model. \textit{Astrobiology}, \textit{6}, 34--47.}

\bigskip

\noindent{\color[rgb]{0.0,0.0,0.039215688}
Tinetti, G., Rashby, S., and Yung, Y. L. (2006b) Detectability of Red-Edge-shifted Vegetation on Terrestrial Planets Orbiting M Stars. \textit{Astrophys. J.}, \textit{644}, L129--L132.}

\bigskip

\noindent{\color[rgb]{0.0,0.0,0.039215688}
Traub, W. and Oppenheimer, B. (2010) Direct Imaging of Exoplanets, In \textit{Exoplanets} (S Seager., ed.), pp. 111-156. University of Arizona Press, Tucson.}

\bigskip

\noindent{\color[rgb]{0.0,0.0,0.039215688}
Tsuji, T. (2002) Dust in the Photospheric Environment: Unified Cloudy Models of M, L, and T Dwarfs. \textit{Astrophys. J.,} \textit{621}, 1033-1048. }

\bigskip

\noindent{\color[rgb]{0.0,0.0,0.039215688}
Tsuji, T., T. Nakajima, \& K. Yanagisawa (2004) Dust in the Photospheric Environment. II. Effect on the Near-Infrared Spectra of L and T Dwarfs. \textit{Astrophys. J.,} \textit{607}, 511-529. }

\bigskip

\noindent{\color[rgb]{0.0,0.0,0.039215688}
Tsuji, T. (2005) Dust in the Photospheric Environment. III. A Fundamental Element in the Characterization of Ultracool Dwarfs. \textit{Astrophys. J.,} \textit{621}, 1033-1048. }

\bigskip

\noindent{\color[rgb]{0.0,0.0,0.039215688}
Valencia, D., O'Connell, R. J., and Sasselov, D. D. (2007) Inevitability of Plate Tectonics on Super-Earths. \textit{Astrophys. J.}, \textit{670}, L45-L48}

\bigskip

\noindent{\color[rgb]{0.0,0.0,0.039215688}
van de Hulst, H. C. (1957) Light Scattering by Small Particles, John Wiley \& Sons, New York.}

\bigskip

\noindent{\color[rgb]{0.0,0.0,0.039215688}
van de Hulst, H. C. (1980) Multiple light scattering. Vols. 1 and 2., Academic Press.}

\bigskip

\noindent{\color[rgb]{0.0,0.0,0.039215688}
van Diedenhoven, B., A.M. Fridlind, A.S. Ackerman, and B. Cairns (2012a) Evaluation of hydrometeor phase and ice properties in cloud-resolving model simulations of tropical deep convection using radiance and polarization measurements, \textit{J. Atmos. Sci.}, \textit{69}, 3290-3314.}

\bigskip

\noindent{\color[rgb]{0.0,0.0,0.039215688}
van Diedenhoven, B., B. Cairns, I.V. Geogdzhayev, A.M. Fridlind, A.S. Ackerman, P. Yang, and B.A. Baum (2012b) Remote sensing of ice crystal asymmetry parameter using multi-directional polarization measurements. Part I: Methodology and evaluation with simulated measurements, \textit{Atmos. Meas. Tech., 5}, 2361-2374.}

\bigskip

\noindent{\color[rgb]{0.0,0.0,0.039215688}
Vasquez, M., Schreier, F., Gimeno Garc\'ia, S., Kitzmann, D., Patzer, A. B. C., Rauer, H., and Trautmann, T. (2013) Infrared Radiative Transfer in Atmospheres of Earth-like Planets around F, G, K, and M-type stars - II. Thermal Emission Spectra In[FB02?]uenced by Clouds. \textit{Astron. \& Astrophys., submitted}}

\bigskip

\noindent{\color[rgb]{0.0,0.0,0.039215688}
Visscher, C., Lodders, K., and Fegley, B. (2006) Atmospheric Chemistry in Giant Planets, Brown Dwarfs, and Low-Mass Dwarf Stars. II. Sulfur and Phosphorus. \textit{Astrophys. J.}, \textit{648}, 1181.}

\bigskip

\noindent{\color[rgb]{0.0,0.0,0.039215688}
Voshchinnikov, N. V., II'in, V. B., Henning, Th., and Dubkova, D. N. (2006) Dust extinction and absorption: the challenge of porous grains. \textit{Astron. \& Astrophys.}, \textit{445}, 167-177.}

\bigskip

\noindent{\color[rgb]{0.0,0.0,0.039215688}
Weidenschilling, S. (1988) Formation processes and time scales for meteorite parent bodies. In \textit{Meteorites and the early solar system} (J.F Kerridge and M.S. Matthews, eds.), pp. 348-371. University of Arizona Press, Tucson.}

\bigskip

\noindent{\color[rgb]{0.0,0.0,0.039215688}
Williams, M. and Loyalka, S. (1991) Aerosol science: theory and practice : with special applications to the nuclear industry. Pergamon Press.}

\bigskip

\noindent{\color[rgb]{0.0,0.0,0.039215688}
Witte, S,., Helling, Ch., and Hauschildt, P. (2009) Dust in brown dwarfs and extra-solar planets. II. Cloud formation for cosmologically evolving abundances. \textit{Astron. \& Astrophys.}, \textit{506}, 1367-1380.}

\bigskip

\noindent{\color[rgb]{0.0,0.0,0.039215688}
Witte, S., Ch. Helling, T. Barman, N. Heidrich, \& P. Hauschildt (2011) Dust in brown dwarfs and extra-solar planets. III. Testing synthetic spectra on observations. \textit{Astron. \& Astrophys.,} \textit{529}, A44 }

\bigskip

\noindent{\color[rgb]{0.0,0.0,0.039215688}
Woitke, P., \& C. Helling (2003) Dust in brown dwarfs. II. The coupled problem of dust formation and sedimentation. \textit{Astron. \& Astrophys.}, \textit{399}, 297-313.}

\bigskip

\noindent{\color[rgb]{0.0,0.0,0.039215688}
Wordsworth, R. D., Forget, F., Selsis, F., et al. (2010) Is Gliese 581d habitable? Some constraints from radiative-convective climate modeling. \textit{Astron. \& Astrophys., 522}, A22.}

\bigskip

\noindent{\color[rgb]{0.0,0.0,0.039215688}
Wordsworth, R. D., Forget, F., Selsis, F., et al. (2011) Gliese 581d is the First Discovered Terrestrial-mass Exoplanet in the Habitable Zone. \textit{Astrophys. J.}, \textit{733}, L48.}

\bigskip

\noindent{\color[rgb]{0.0,0.0,0.039215688}
Wright , E. L. (1987) Long-wavelength absorption by fractal dust grains. \textit{Astrophys. J.}, \textit{320}, 818-824.}

\bigskip

\noindent{\color[rgb]{0.0,0.0,0.039215688}
Zahnle, K., M.S. Marley, \& J.J. Fortney (2009) Thermometric Soots on Warm Jupiters? arXiv:0911.0728 }

\bigskip

\noindent{\color[rgb]{0.0,0.0,0.039215688}
Zsom, A., C. W. Ormel, C. Guettler, J. Blum, and C. P. Dullemond (2010) The outcome of protoplanetary dust growth: pebbles, boulders, or planetesimals? II. Introducing the bouncing barrier; \textit{Astron. \& Astrophys., 513}, A57}

\bigskip

\noindent{\color[rgb]{0.0,0.0,0.039215688}
Zsom, A., Kaltenegger, L., and Goldblatt, C. (2012). A 1D microphysical cloud model for Earth, and Earth-like exoplanets: Liquid water and water ice clouds in the convective troposphere. \textit{Icarus}, \textit{221}, 603--616.}

\end{document}